\documentclass[hyper]{JHEP3}
\usepackage{graphicx,amssymb,bm,latexsym,amsmath}
\usepackage{epstopdf}
\usepackage{caption}
\usepackage{subfigure}
\usepackage[countmax]{subfloat}
\usepackage[toc,page]{appendix}
\newcommand{\bej}[1]{ \begin{equation}\label{#1} }
\newcommand{\eej}{\end{equation}}
\newcommand{\beaj}[1]{\begin{eqnarray}\label{#1} }
\newcommand{\eeaj}{\end{eqnarray}}
\newcommand{\eq}[1]{(\ref{#1})}
\def\ZZZ{{\hskip-3pt\hbox{ Z\kern-1.6mm Z}}}
\def\zzz{{\hskip-3pt\hbox{ z\kern-1mm z}}}

\newcommand{\bd}{\bar{\rm D}}

\newcommand{\N}{\frac{m_{2}}{k_{2}}-\frac{m_{1}}{k_{1}}}

\newcommand{\be}{\begin{equation}}
\newcommand{\ee}{\end{equation}}
\newcommand{\ben}{\begin{eqnarray}\displaystyle}
\newcommand{\een}{\end{eqnarray}}

\def\one{{\hbox{ 1\kern-.8mm l}}}
\def\zero{{\hbox{ 0\kern-1.5mm 0}}}

\def\be{\begin{equation}}       
\def\ee{\end{equation}}         

\def\bea{\begin{eqnarray}}      
\def\eea{\end{eqnarray}}
                                
\def\ba{\begin{array}}
\def\ea{\end{array}}
\def\bd{\begin{displaymath}}
\def\ed{\end{displaymath}}

\def\eq{\begin{equation}}
\def\eqe{\end{equation}}
\def\eqa{\begin{eqnarray}}
\def\eqae{\end{eqnarray}}
\def\ena{\end{eqnarray}}

\def\unit{1 \hskip-.3em \raise2pt\hbox{$ \scriptstyle |$ } }

\def\bd{\begin{displaymath}}
\def\ed{\end{displaymath}}

\def\6{\partial}
\def\N4{{\cal N}=4}

\def\bop#1{\setbox0=\hbox{$#1M$}\mkern1.5mu
        \vbox{\hrule height0pt depth.04\ht0
        \hbox{\vrule width.04\ht0 height.9\ht0 \kern.9\ht0
        \vrule width.04\ht0}\hrule height.04\ht0}\mkern1.5mu}

\def\>{\rangle} 

\def\<{\langle} 
\def\Dsl{D \hskip-.6em \raise1pt\hbox{$ / $ } }
\def\to{\rightarrow}

\def\+{\oplus}

\def\as2{AdS_3\times S^3_1 \times S^3_2}

\title{Pulsating strings in I-brane background}

\author{Sagar Biswas\\
Department of Physics, Ramakrishna Mission Vidyamandira, \\ Belur Math, Howrah 711 202, India

Email: \email{biswas.sagar09iitkgp@gmail.com}}
\author{Nibedita Padhi \\
	Department of Physics, Indian Institute of Technology Kharagpur,\\  Kharagpur 721 302, India

	Email: \email{nibedita.phy@iitkgp.ac.in}}
\author{Kamal L Panigrahi \\
 Department of Physics, Indian Institute of Technology Kharagpur,\\  Kharagpur 721 302, India

Email: \email{panigrahi@phy.iitkgp.ac.in}}
\abstract{We study pulsating strings both on a single stack of NS5-branes and two orthogonal stacks of NS5-branes (the so called I-brane) by using the Polyakov form of the fundamental string action. For the I-brane background, by using a symmetry that decouples the two spheres from the flat geometry, we study pulsating solutions of the string when it pulsates  on both the  spheres independently and simultaneously. We finally derive the energy of pulsating strings as a function of adiabatic invariant oscillation number in these cases and studied some limiting cases in detail.}

\keywords{Bosonic strings, AdS/CFT correspondence}

\begin{document}
\section{Introduction}
After its appearance $AdS/CFT$ correspondence \cite{a} has turned out to be one of the most powerful tools of modern theoretical physics. The most successful version of the correspondence relies on the duality between type IIB string states in $AdS_5\times S^5$ and certain operators in a four dimensional Conformal Field Theory (CFT) living on the boundary of $AdS_5$. But even for the very well studied  $\mathcal{N}=4$ Supersymmetric Yang-Mills (SYM) theory in four dimensions and dual type IIB superstring in the compactified $AdS_5$ space, finding an exact matching between string states and dual operators from both sides is very difficult. One can check the correspondence beyond the supergravity approximation by studying various classical string solutions in different backgrounds and using the dispersion relation of such strings in the large charge limit, one can look for boundary operators dual to them. The development of the integrability technique \cite{Minahan:2002ve}-\cite{Bena:2003wd} associated with both sides of the duality has reduced the problem of solving the spectra in large charge limit to the problem of solving a set of algebraic Bethe equations. Precisely, the integrability has improved the understanding of the equivalence between the Bethe equation of the spin chain and the corresponding realization of the worldsheet symmetries of the classical $AdS_5 \times S^5$ string sigma model \cite{Kazakov:2004qf},\cite{Zarembo:2004hp}. The corresponding Bethe equations are based on the knowledge of the S-matrix which focusses on the scattering of world-sheet excitations of the gauge-fixed string sigma model, or equivalently, the excitations of a certain spin chain in the dual gauge theory \cite{Beisert:2004hm}-\cite{Beisert:2005tm} . 
 
\medskip

Over the years, a large variety of rotating and spinning strings has been studied in backgrounds arising from the $AdS_5 \times S^5$ string sigma model. This include well known solutions like folded spinning strings \cite{Gubser:2002tv}, spiky strings \cite{Kruczenski:2004wg} and giant magnons \cite{Hofman:2006xt} and their dual gauge theories have been analysed in great details. Pulsating strings were first introduced in \cite{Minahan:2002rc} and are expected to be dual to highly excited sigma model operators. For example the most general pulsating string in $S^5$ charged under the isometry group $SO(6)$ will have a dual operator of the form Tr($X^{J_1}Y^{J_2}Z^{J_3}$), where $X, Y$ and $Z$ are the chiral scalars and $J_i$'s are the R-charges from the SYM theory. Pulsating string solutions have better stability than non-pulsating ones \cite{Khan:2005fc} and they are time-dependent as opposed to the usual rigidly rotating string solution. Pulsating strings have been thoroughly generalized in \cite{Khan:2003sm}-\cite{Kruczenski:2004cn}, and have been studied in a number of backgrounds having varying degrees of supersymmetry \cite{Chen:2008qq}-\cite{Arnaudov:2010by}.
 A string rotating and at the same time oscillating have been derived in \cite{Park:2005kt}, and the generalization of this was done with extra angular momenta in \cite{Pradhan:2013sja}. Various other developments and explorations in different backgrounds have been done by different authors \cite{Beccaria:2010zn}-\cite{Banerjee:2016xbb}.
\medskip

In the present paper we will study pulsating string solutions in NS5-brane and I-brane background. In string theory NS5-branes are interesting on their own right as in the near horizon limit the theory on the worldvolume correspond to the little string theory (LST) \cite{Aharony:1999ks},\cite{Kutasov:2001uf}. LST is a nonlocal field theory which has not been understood properly until now, hence it would be a good exercise to analyze the solution in various limits and if possible find out some operators in dual field theory. It has been found that in the near horizon limit the world sheet theory of NS5-brane is exactly solvable, so from the bulk theory point of view the theory is integrable. But, only a very little is known about the boundary theory, hence from that perspective it is rather hard to make definite satements about the exact nature of the theory. We will also study the pulsating string solutions in the I-brane background which arises from the 1+1 dimensional intersection of two orthogonal stacks of NS5-branes, with one set of branes lying along $(x^0, x^1,\cdots, x^5)$, and other set lying along $(x^0, x^6,\cdots, x^9)$ directions \cite{Itzhaki:2005tu}. In the near horizon geometry when all five branes are coincident the S-dual picture is given by
$$R^{2,1} \times R_{\phi} \times SU(2)_{k_1} \times SU(2)_{k_2}$$
where $R_{\phi}$ is one combination of the radial directions away from the two sets of NS5-branes, and $R^{2,1}$ (whose coordinates are $x^0, x^1$) are another combination of two radial directions. The two $SU(2)$s with levels $k_{1,2}$ describe the angular three-spheres corresponding to $(R^4)_{2345}$ and $(R^4)_{6789}$. As mentioned in \cite{Itzhaki:2005tu} this background exhibits a higher Poincare symmetry, ISO(2,1), than the expected ISO(1,1) and twice as many supercharges one might expect. 
\medskip

Various aspects and semiclassical strings on NS5-brane and I-brane have been studied by probing the geometry with both fundamental strings and D1-strings in \cite{Kluson:2007qu}-\cite{Biswas:2023lkj}.
Very recently we have studied rigidly rotating string in I-brane using Polyakov action of string \cite{Biswas:2023lkj} and we find Polyakov action of string completely decouple the two spheres. This allows us to study the rotating string on both the spheres simultaneously and independently and we find two sets of giant magnon relations for a particular set of values of the parameters while two sets of single spike relations for all other set of values of the parameters. To see whether this type of behaviour also exists for pulsating string solutions, we probe fundamental string in I-brane background with Polyakov action of string. As I-brane arises from two orthogonal stacks of NS5-branes we are expecting that we will obtain two independent class of pulsating string solutions simultaneously on both the spheres. To compare with various relations we also studied pulsating string solutions in NS5-brane background using the same action of strings. The rest of the paper is organized as follows. In section \eqref{sec2}, we will briefly review the near-horizon geometry of NS5-brane and solve the equations of motion for the pulsating strings which are consistent with Virasoro constraints. We also solve for the string profile and find the oscillation numbers alongwith the energy-oscillation number relation. In section \eqref{sec3}, after reviewing the I-brane background briefly we solve the string equations of motion and show that in this case the string can pulsate independently and simultaneously on both the spheres. We also find the string profiles, oscillation numbers and energy-oscillation number relation when energy is equally and unequally distributed among the spheres of I-brane. In section \eqref{sec4} we will summarize our results and comment on further investigations.

\section{Pulsating strings in NS5-brane}\label{sec2}
The classical solution of $N$ NS5-brane is given by the following form of metric, NS-NS two form field and dilaton,
\begin{eqnarray}
    ds^2 &=& -dt^2 + \sum_{i=5}^9 dx_i^2 + H(r) (dr^2 + r^2(d\theta^2 + \sin^2\theta d\phi^2 + \cos^2\theta d\psi^2)) \ , \nonumber \\ B_{\phi\psi} &=& 2N\sin^2\theta d\phi \wedge d\psi \ , ~~~ e^{2(\Phi - \Phi_0)} = H(r) \ , ~~~ H(r) = 1 + \frac{Nl_s^2}{r^2} \ ,
\end{eqnarray}
where $x^i, i=5,\cdots,9$ labels the world-volume directions of NS5-brane, $H(r)$ is the Harmonic function in the transverse space of the NS5-branes and $l_s$ is the string length. In the near horizon limit, $r\to 0$, one can ignore $1$ in the $H(r)$, and the solution would look like,
\begin{eqnarray}
    ds^2 &=& -dt^2 + \sum_{i=5}^9 dx_i^2 + Nl_s^2 \Big(\frac{dr^2}{r^2} + d\theta^2 + \sin^2\theta d\phi^2 + \cos^2\theta d\psi^2\Big) \ , \nonumber \\ b_{\phi\psi} &=& 2N\sin^2\theta \ , ~~~ e^{2(\Phi - \Phi_0)} = \frac{Nl_s^2}{r^2} \ ,
\end{eqnarray}  
where $B_{\phi\psi}=b_{\phi\psi} d\phi \wedge d\psi$. To proceed further we will rescale $t \to \sqrt{N}l_st, x_i \to \sqrt{N}l_s x_i$ and introduce the variable $\rho$ that is related to $r$ as, 
\begin{equation}
    \rho = \ln \Big(\frac{r}{\sqrt{N}l_s}\Big) \ ,
\end{equation}
and setting $l_s=1$. Under all this operation the final form of the metric and NS-NS two form field becomes,
\begin{eqnarray}
    ds^2 &=& N(-dt^2 + \sum_{i=5}^9 dx_i^2 + d\rho^2 + d\theta^2 + \sin^2\theta d\phi^2 + \cos^2\theta d\psi^2\Big) \ , \nonumber \\ b_{\phi\psi} &=& 2N\sin^2\theta \ . \label{metric}
\end{eqnarray}
In this section we will restrict to the background metric and NS-NS two form field given by (\ref{metric}).

\subsection{String equations of motion}
To study a fundamental string coupled to NS-NS B-field, we use the Polyakov action with a WZ term,
\begin{eqnarray}
S=-\frac{\sqrt{\lambda}}{4\pi}\int d\sigma d\tau
[\sqrt{-\gamma}\gamma^{\alpha \beta}g_{MN}\partial_{\alpha} X^M
\partial_{\beta}X^N - \epsilon^{\alpha \beta}\partial_{\alpha} X^M
\partial_{\beta}X^N b_{MN}] \ ,
\end{eqnarray}
where $\lambda$ is the 't Hooft coupling, $\gamma^{\alpha \beta}$ is the worldsheet metric and $\epsilon^{\alpha \beta}$ is the antisymmetric tensor defined as $\epsilon^{\tau \sigma}=-\epsilon^{\sigma \tau}=1$.

Variation of the action with respect to $X^M$ gives us the following equations of motion
\begin{eqnarray}
2\partial_{\alpha}(\eta^{\alpha \beta} \partial_{\beta}X^Ng_{KN})
&-& \eta^{\alpha \beta} \partial_{\alpha} X^M \partial_{\beta}
X^N\partial_K g_{MN} - 2\partial_{\alpha}(\epsilon^{\alpha \beta}
\partial_{\beta}X^N b_{KN}) \nonumber \\ &+& \epsilon ^{\alpha \beta}
\partial_{\alpha} X^M \partial_{\beta} X^N\partial_K b_{MN}=0 \ ,
\end{eqnarray}
and variation with respect to the metric gives the two Virasoro
constraints,
\begin{eqnarray}
g_{MN}(\partial_{\tau}X^M \partial_{\tau}X^N +
\partial_{\sigma}X^M \partial_{\sigma}X^N)&=&0 \ ,\label{v1} \\ 
g_{MN}(\partial_{\tau}X^M \partial_{\sigma}X^N)&=&0 \label{v2}\ .
\end{eqnarray}
We use the conformal gauge (i.e.
$\sqrt{-\gamma}\gamma^{\alpha \beta}=\eta^{\alpha \beta}$) with
$\eta^{\tau \tau}=-1$, $\eta^{\sigma \sigma}=1$ and $\eta^{\tau
\sigma}=\eta^{\sigma \tau}=0$) to solve the equations of motion.
\medskip

For studying a generic class of pulsating strings we use the ansatz,
\begin{eqnarray}
    && t = t(\tau) = \kappa\tau  \ , ~~~ x_i = v_i\sigma \ , ~~~ \rho = \mu\sigma \ , ~~~ \theta=\theta(\tau) \ , ~~~ \phi = m\sigma \ , ~~~  \psi = \psi(\tau) \ . 
\end{eqnarray}
Solving $t$ and $\psi$ equations we obtain,
\begin{eqnarray}
    \dot{t} &=& \kappa \ , \nonumber \\ \dot{\psi} &=& \frac{c_1 + 2m \sin^2\theta}{\cos^2\theta} \ . 
\end{eqnarray}
The equations for $x_i$, $\rho,$ and $\phi$ satisfies trivially. The other non-trivial contribution comes from $\theta$ equation, which gives,
\begin{eqnarray}
    \ddot{\theta} &=& \Big[3m^2 - \frac{(c_1+2m)^2}{\cos^4\theta}\Big] \sin\theta\cos\theta \ . 
\end{eqnarray}
Integrating the above equation we get,
\begin{eqnarray}
    \dot{\theta}^2 &=& c_2 + 3m^2\sin^2\theta - \frac{(c_1+2m)^2}{\cos^2\theta}  \label{th} \ ,
\end{eqnarray}
where $c_{1,2}$ are the integrating constants.
\medskip

Now, the Virasoro constraint $g_{MN}(\partial_{\tau}X^M\partial_{\tau}X^N + \partial_{\sigma}X^M\partial_{\sigma}X^N)=0$, gives 
\begin{equation}
    \dot{\theta}^2  = \alpha^2 + 4m^2 + 4c_1m + 3m^2\sin^2\theta - \frac{(c_1+2m)^2}{\cos^2\theta}  \label{Vir} \ ,
\end{equation}
where $\alpha^2 = \kappa^2 - \mu^2 - \sum v_i^2$. The other Virasoro $g_{MN}\partial_{\tau}X^M\partial_{\sigma}X^N=0$ is trivially satisfied. 
Comparing the Virasoro constraint (\ref{Vir}) with the equation of motion (\ref{th}), we get the following relation between various constants,
\begin{equation}
    c_2 = \alpha^2 + 4m^2 + 4c_1m \label{conda} \ .
\end{equation}
\medskip

To determine the conserved charges associated to the string motion, we start from the full form of the sigma model action in background \eqref{metric},
\begin{eqnarray}
   S = -\frac{\sqrt{\lambda}}{4\pi} \int d\tau d\sigma \Big[ \dot{t}^2 + \sum_{i=5}^9 v_i^2 + \mu^2  - \dot{\theta}^2  + m^2\sin^2\theta  -\dot{\psi}^2\cos^2\theta + 4m\dot{\psi}\sin^2\theta \Big] \ .
\end{eqnarray}
From this action we can easily determine the conserved charges using the Noether procedure, and they can be written as,
\begin{eqnarray}
    E &=& - \int \frac{\partial\mathcal{L}}{\partial \dot{t}} ~d\sigma = \sqrt{\lambda}\kappa \ , \nonumber \\
J &=&  \int \frac{\partial\mathcal{L}}{\partial \dot{\psi}} ~d\sigma = \sqrt{\lambda} c_1  \ . 
\end{eqnarray}
Here $E$ is the energy and $J$ is the angular momenta on the sphere. Rescaling them as,
\begin{eqnarray}
    \mathcal{E} = \frac{E}{\sqrt{\lambda}} = \kappa \ , ~~~ \mathcal{J} = \frac{J}{\sqrt{\lambda}} = c_1 \ .
\end{eqnarray}
Now expressing $\kappa$ and $c_1$ in terms of these conserved charges and for simplicity using $\sum v_i^2 = 1$, we can express the constraint equation (\ref{conda}) as,
\begin{equation}
    c_2 = \mathcal{E}^2 - 1 - \mu^2 + 4m^2 + 4m\mathcal{J} \ . \label{condb}
\end{equation}

\subsection{String profile}
From (\ref{th}) we find as $\theta$ varies from $0$ to $\pi/2$, $\dot{\theta}^2$ varies from $\mathcal{E}^2 - 1 - \mu^2 - \mathcal{J}^2$ to infinity. This looks like the equation of motion of a particle moving in an effective potential $V(\theta)$, where $\theta$ varies from a minimal to a maximal value. Note that the $\theta$ equation can be written in the form,
\begin{equation}
    \ddot{x} + 3m^2[-(R_{-}+R_{+})x + 2x^3]=0 \ , \label{tha}
\end{equation}
where $x=\sin\theta$. Equation (\ref{tha}) can be compared with Duffing oscillator equation ($\ddot{x}+\gamma \dot{x} + \omega_0^2 x + \epsilon x^3 = f_0 \cos(\omega t)$) without damping and driving force if $(R_{-}+R_{+})$ is negative. Integrating (\ref{tha}), we get,
\begin{eqnarray}
    \dot{x}^2 &=& 3m^2x^2(1-x^2) + c_2(1-x^2) - (\mathcal{J} + 2m)^2 \nonumber \\ && = 3m^2(x^2 - R_{-})( R_{+}-x^2) \ , \label{thb}
\end{eqnarray}
where 
\begin{eqnarray}
    R_{\pm} &=& \frac{-(c_2 - 3m^2) \pm \sqrt{(c_2 + 3m^2)^2 - 12 m^2 (\mathcal{J} + 2m)^2}}{6m^2}  \ .
\end{eqnarray}

 With proper scaling of the variables, we can write the solution of (\ref{thb}) in terms of standard Jacobi elliptic function \footnote{In the notation we follow $sd(z|m)$ is the solution of $w''(z)^2+w(z)(2m(1-m)w^2(z)-2m+1)=0.$}, provided the initial condition $x(0)=0$:
\begin{eqnarray}
    \sin \theta(\tau) &=& \sqrt{\frac{-R_{+}R_{-}}{R_{+} -R_{-}}}sd\Big(\sqrt{3(R_{+} -R_{-})}m\tau ~~,~~\sqrt{\frac{R_{+}}{R_{+}-R_{-}}} \Big) \label{profile1}
\end{eqnarray}
Using the property of Jacobi functions that $sd(z|m)=sd(z+4\mathbb{K}(m)|m)$ and as usual taking only the real period, we can find the condition for time-periodic solution for $\theta$ is,
\begin{equation}
 0 < \frac{R_{+}}{R_{+}-R_{-}} < 1
\end{equation}
 This translate to the following inequality
\begin{equation}
	c_2 > (\mathcal{J}+2m)^2
\end{equation}
which gives a constraint on the conserved charges so that the string has a pulsating motion. Using this inequality in (\ref{condb}) we get,
   \begin{equation}
   	\mathcal{E}^2 - 1 - \mu^2 - \mathcal{J}^2 > 0 \ . \label{condd}
   \end{equation}
This condition (\ref{condd}) is in tune with our earlier observation about the limits of the oscillation in the $\dot{\theta}^2$ equations.
\medskip  
	\begin{figure}[h!]
	\centering
	\subfigure{\includegraphics[width=0.25\linewidth]{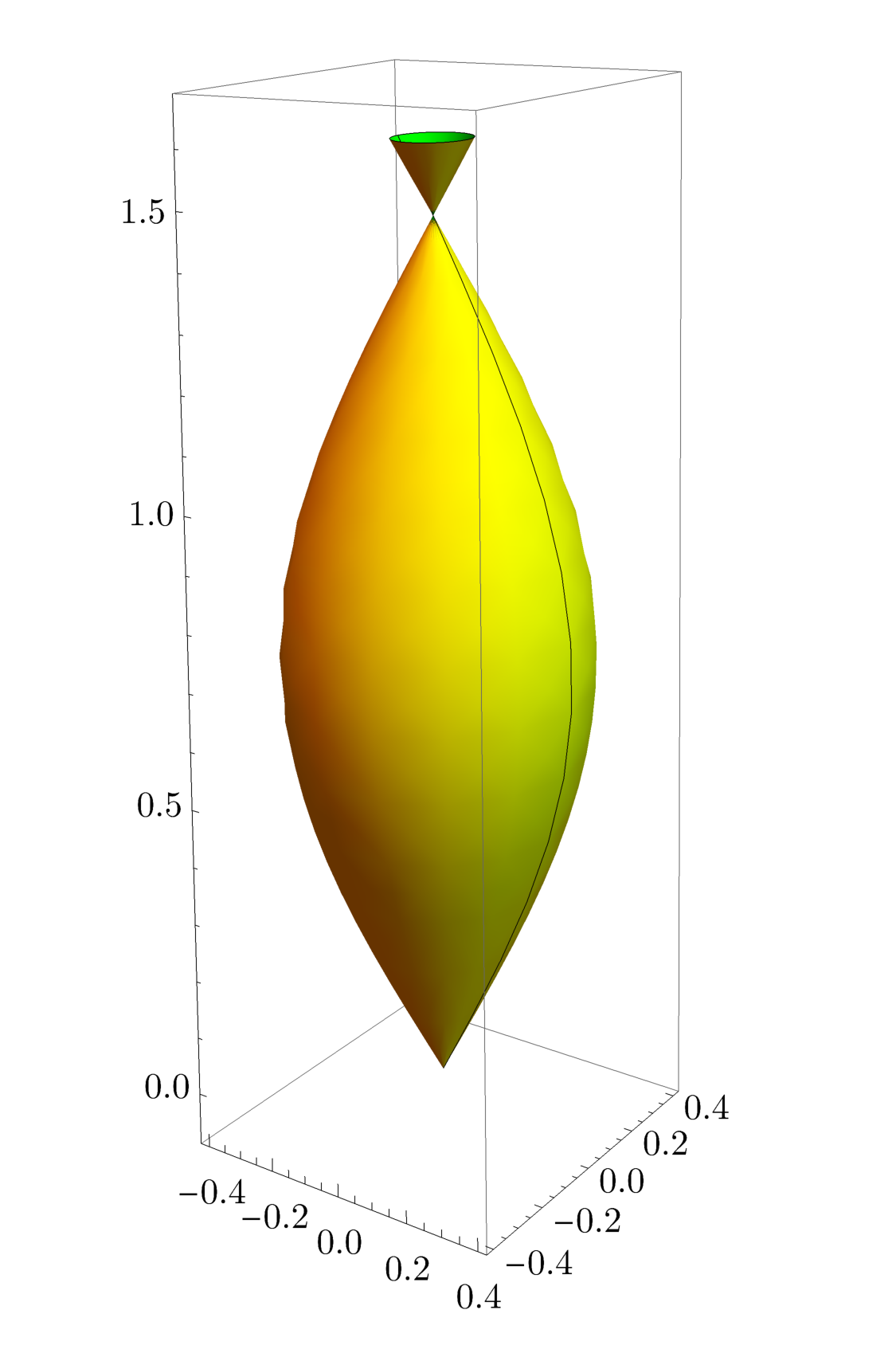}} \hfill
	\subfigure{\includegraphics[width=0.25\linewidth]{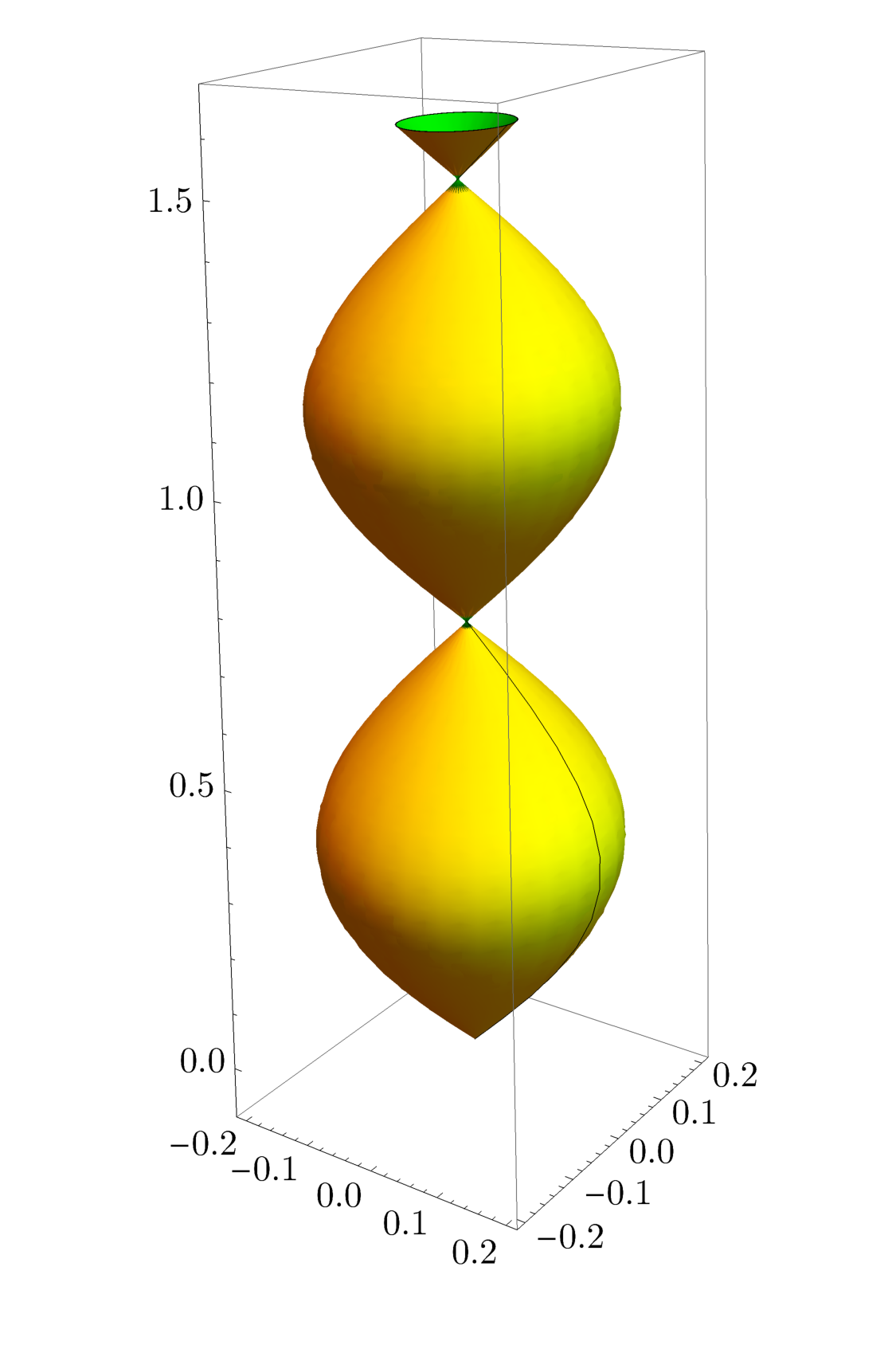}} \hfill
	\subfigure{\includegraphics[width=0.25\linewidth]{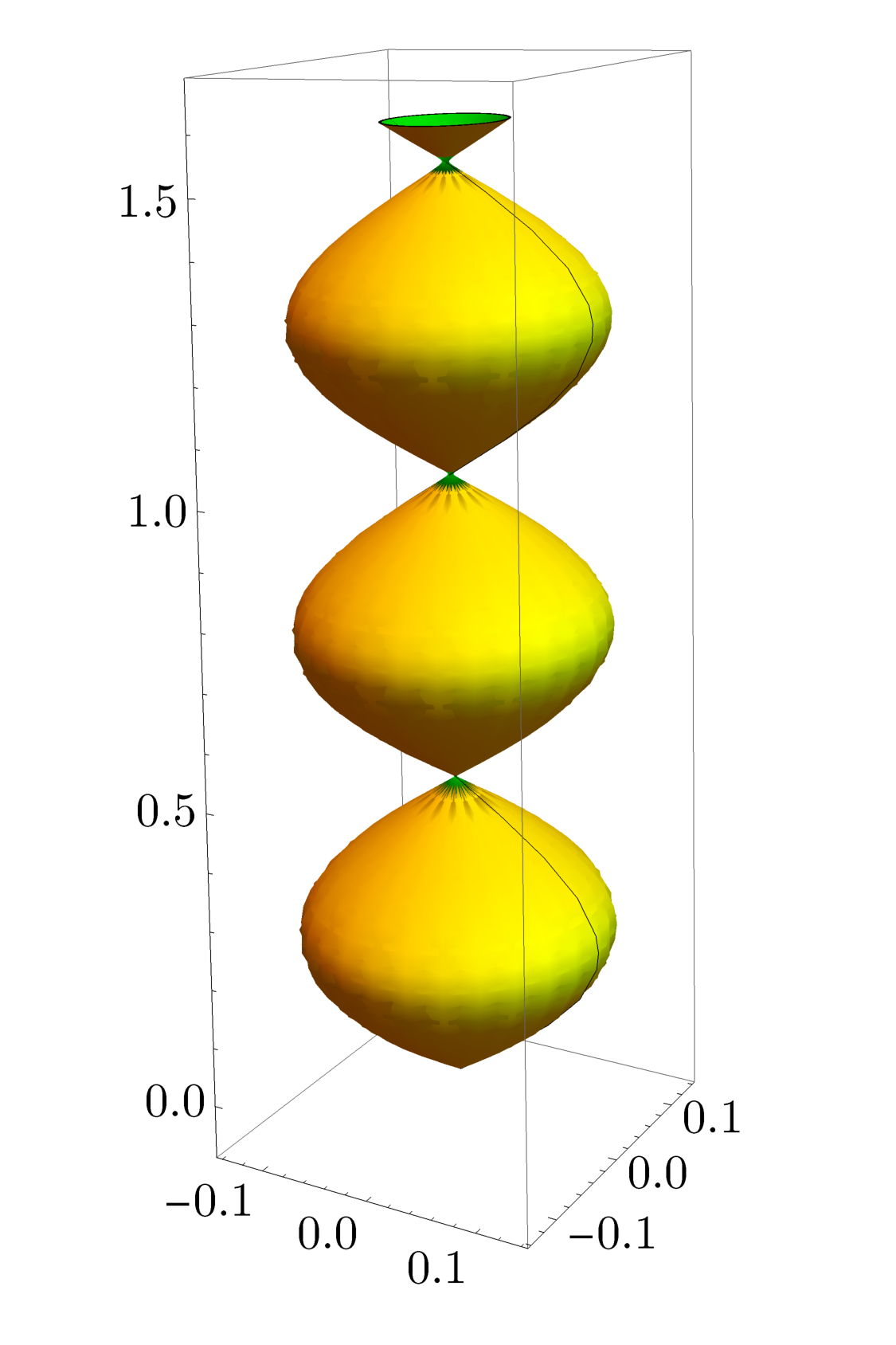}
	}
		\caption{Pulsating string profiles with $\mathcal{E} =1.7 $ and (a)  $m=1$  (b)$m=3$  (c)$m=5$.} \label{fig 1}
\end{figure}

	\begin{figure}[h!]
	\centering
	\subfigure{\includegraphics[width=0.25\linewidth]{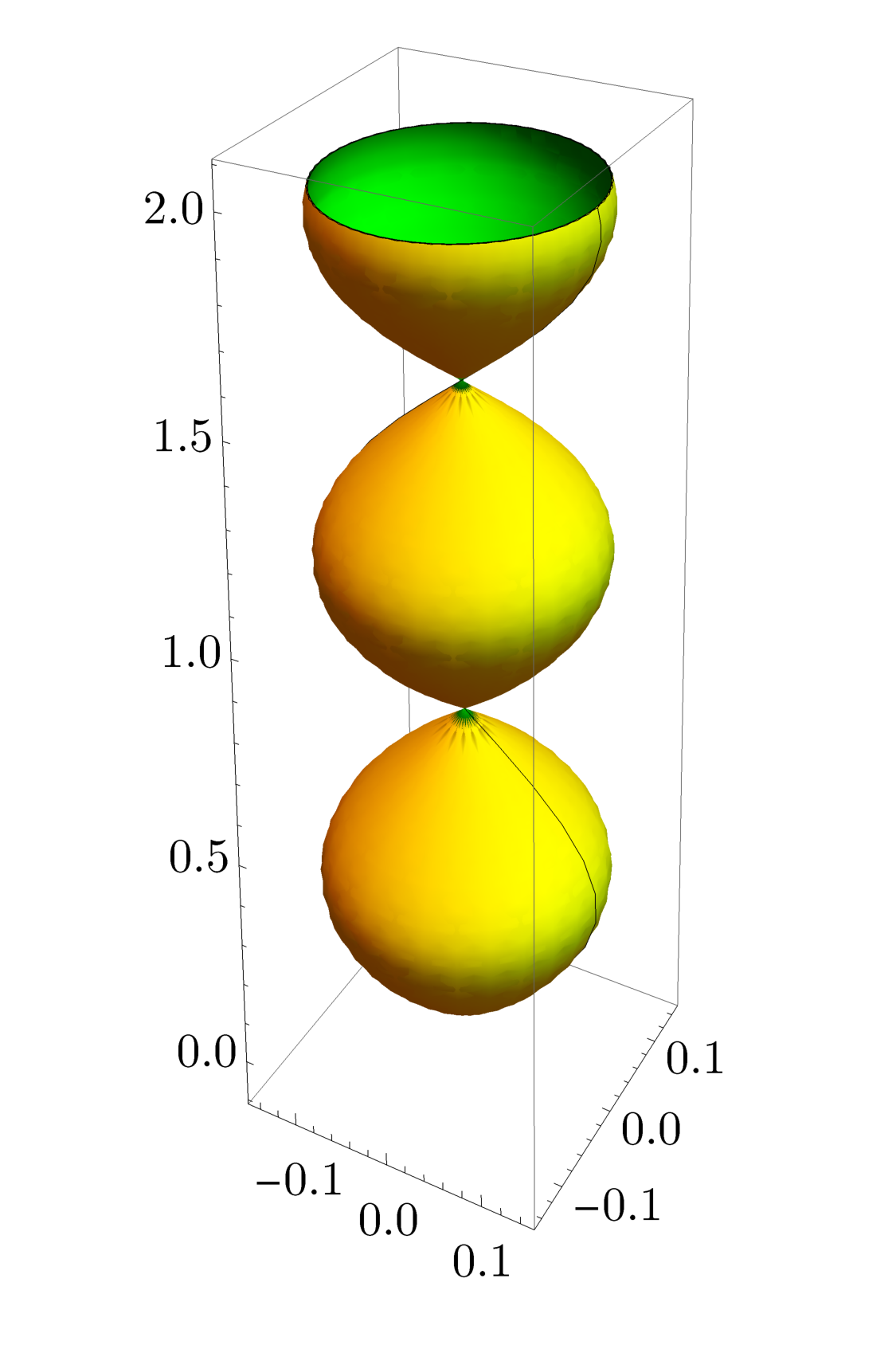}} \hfill
	\subfigure{\includegraphics[width=0.25\linewidth]{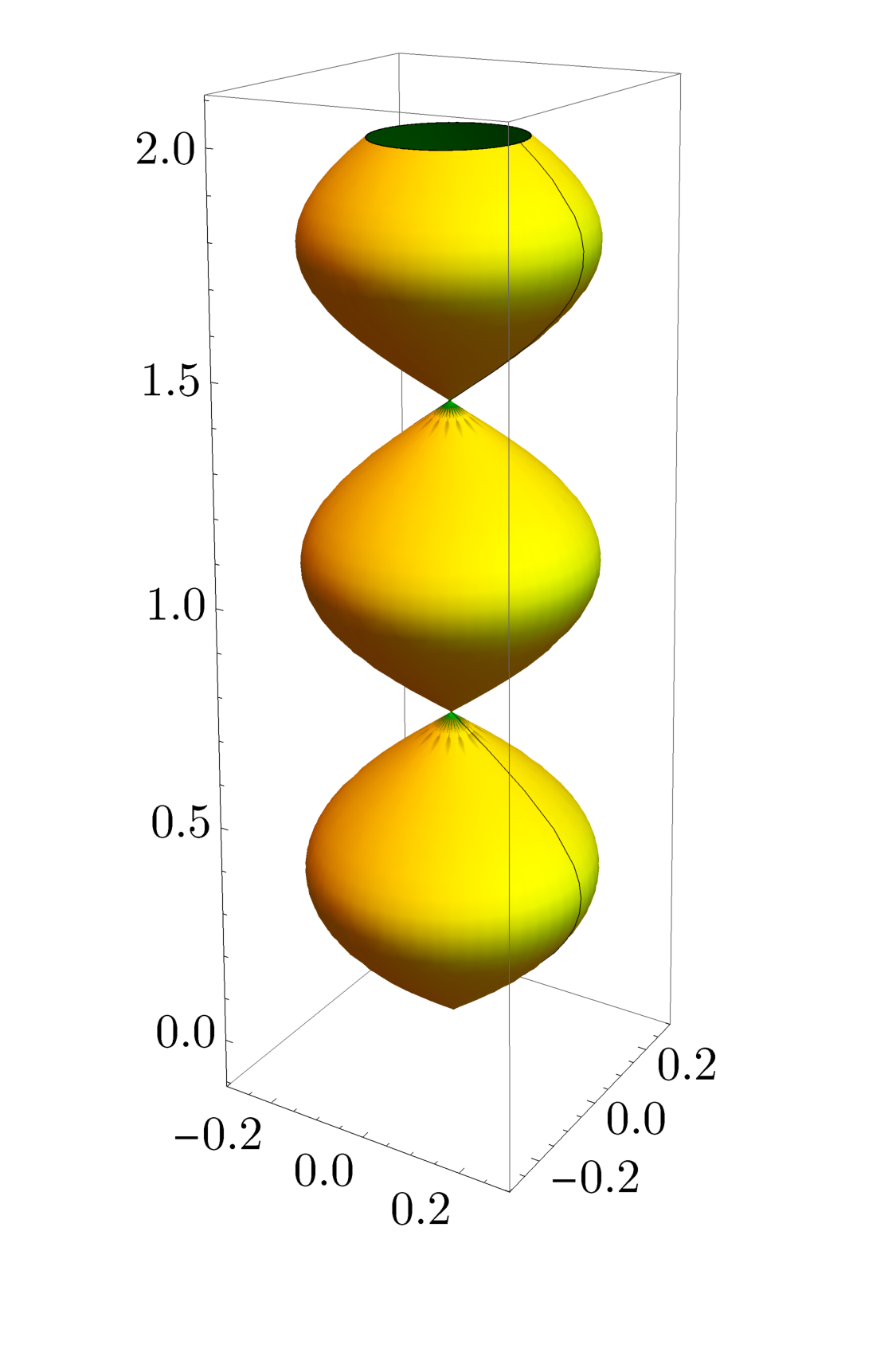}} \hfill
	\subfigure{\includegraphics[width=0.25\linewidth]{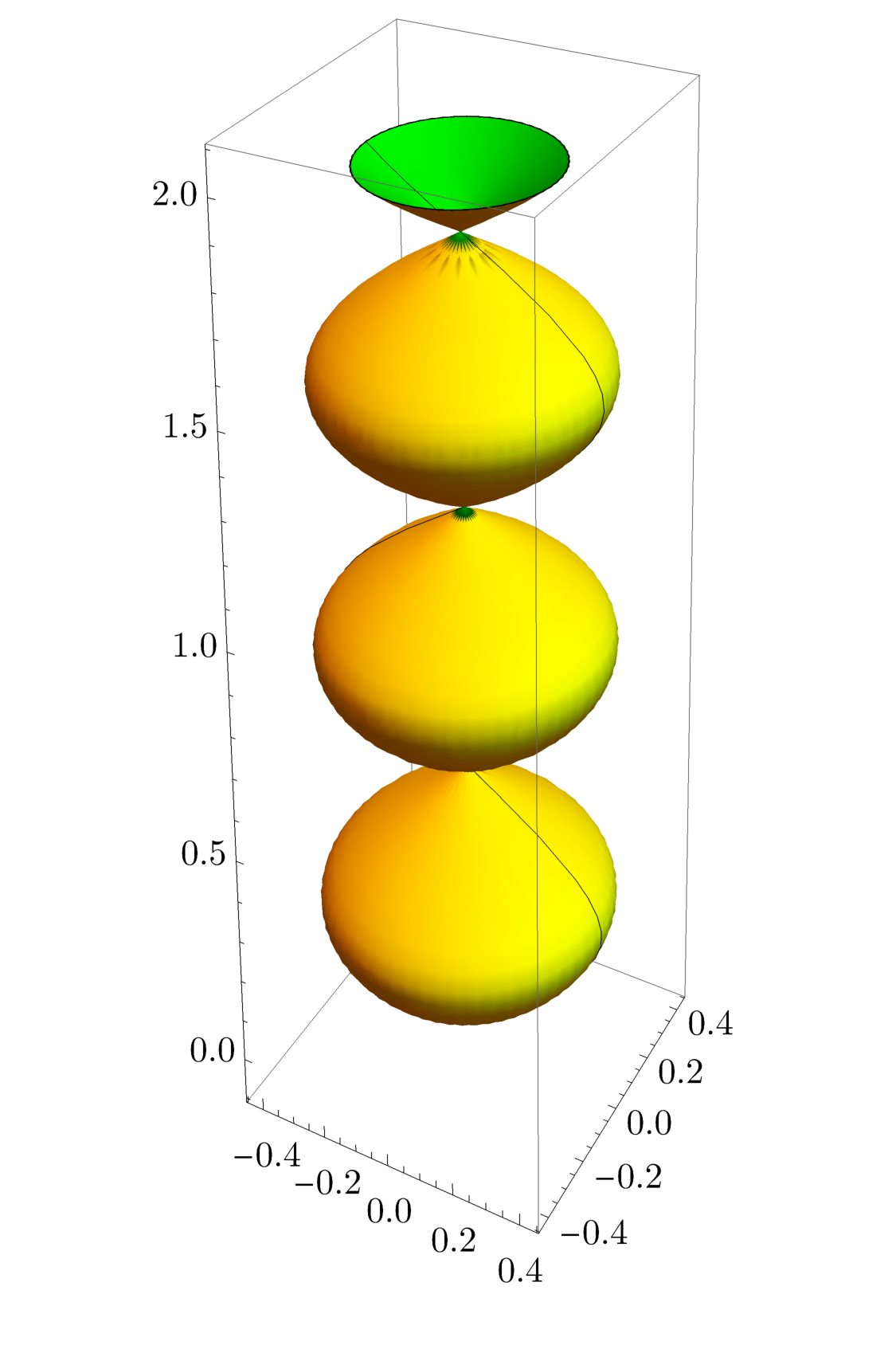}
	}
\caption{Pulsating string profiles with $m =3$ and (a)  $\mathcal{E}=1.6$  (b)$\mathcal{E}=1.9$  (c)$\mathcal{E}=2.5$.} \label{fig 2}
\end{figure}

\begin{figure}[h!]
	\centering
	\includegraphics[width=3.5cm]{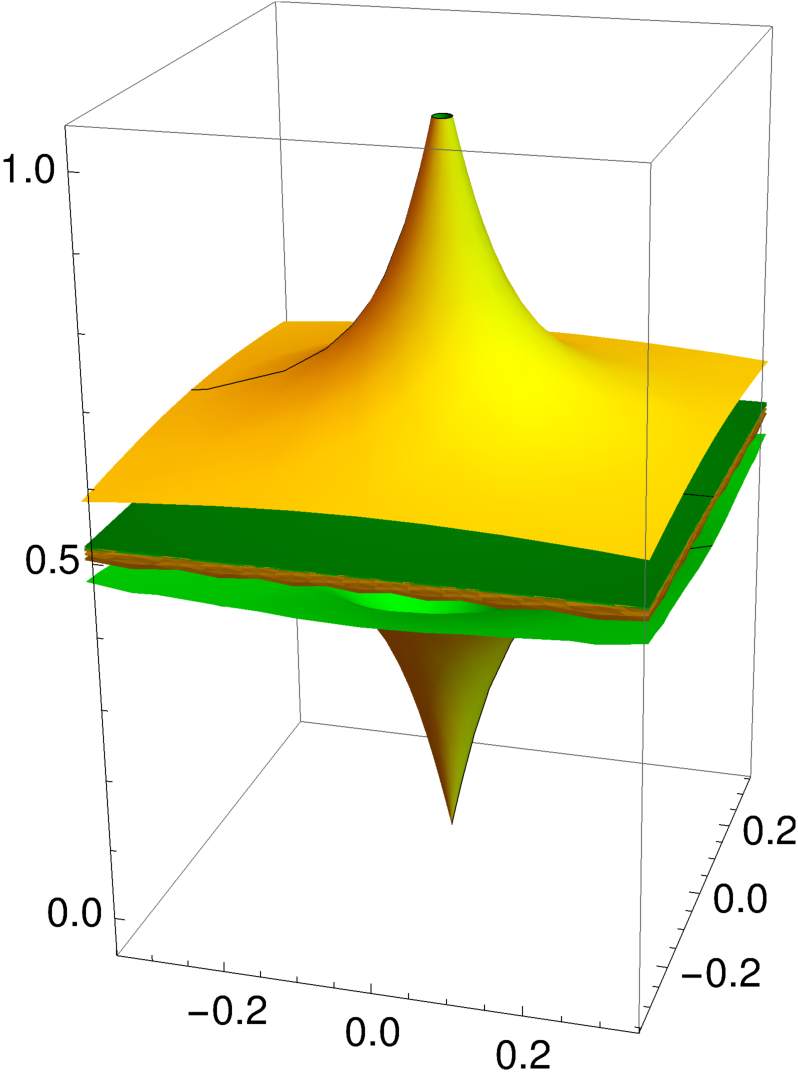}
	\caption{Plot shows unphysical structure when the inequality condition is not obeyed by the solution. } \label{fig 3}
\end{figure}
	
Figure (\ref{fig 1}) shows the string profile in NS5-brane for different values of winding number ($m$) while energy ($\mathcal{E}$), angular momenta ($\mathcal{J}$) and other winding number ($\mu$) remaining constant. These figures clearly indicate that with the increase of $m$, number of lobes increases within a certain value of worldsheet time $\tau$. From these plots we can guess the number of lobes = $(m+1)/2$. We can also see with the increase in the number of lobes the amplitude decreases for a given value of energy. Figure (\ref{fig 2}) shows the plot of string profile with different values of energies while keeping angular momenta and winding numbers ($m$ and $\mu$) fixed. These plots clearly indicate the amplitude of the string profile increases with the increase of energy. We can also note that although we keep $m=3$ fixed but the number of lobes increases with energy within a certain value of the worldsheet time. Hence, number of lobes depends on both the energy and the winding number $m$. String profiles are not sensitive to angular momenta and other winding number $\mu$ unless they take such values that the inequality (\ref{condd}) breaks down, this leads to unphysical structure as the string profile breaks down as shown in figure (\ref{fig 3}). 
\medskip

We can also find the dynamics of the string along the $\psi$ direction by integrating $\frac{d\psi}{d\theta}$ which have the form,
\begin{eqnarray}
    \frac{d\psi}{d\theta} &=& \frac{\mathcal{J} + 2m\sin^2\theta}{\sqrt{3}m\cos\theta \sqrt{(\sin^2\theta - R_{-})( R_{+}-\sin^2\theta )}} \ .
\end{eqnarray}
These can be integrated to find $\psi$ in terms of standard elliptic integrals,
\begin{eqnarray}
    \psi(\tau) &=& \frac{-1}{\sqrt{-3R\textbf{}_{-}}} \Big[\frac{\mathcal{J} + 2m}{m} ~ \Pi \Big(R_{+}, \arcsin\Big(\frac{\sin\theta(\tau)}{\sqrt{R_{+}}}\Big), \frac{R_{+}}{R_{-}}\Big) \nonumber \\ && + 2F \Big( \arcsin\Big(\frac{\sin\theta(\tau)}{\sqrt{R_{+}}}\Big), \frac{R_{+}}{R_{-}} \Big) \Big] \ , 
\end{eqnarray}
where $F(\varphi,k)$ and $\Pi(n,\varphi,k)$ are the incomplete elliptic integrals of first kind and third kind respectively.

\subsection{Oscillation Number}
Now we will use Bohr-Sommerfeld like quantization procedure for the pulsating strings in this background. The oscillation numbers can be written using the canonical momenta conjugate to $\theta$ as follows,
\begin{eqnarray}
    N &=& \sqrt{\lambda}\mathcal{N} = \frac{\sqrt{\lambda}}{2\pi} \oint d\theta  ~ \Pi_{\theta} \nonumber \\ && = \frac{\sqrt{\lambda}}{2\pi} \oint d\theta \sqrt{c_2 + 3m^2\sin^2\theta - \frac{(\mathcal{J} + 2m)^2}{\cos^2\theta}} \ .
\end{eqnarray}
Taking $\sin\theta=x$ we can choose the proper limits and transform the above integral to,
\begin{eqnarray}
    \mathcal{N} &=& \frac{2\sqrt{3}m}{\pi} \int_0^{\sqrt{R_{+}}} \frac{\sqrt{(x^2 - R_{-})( R_{+}-x^2)}}{1-x^2} dx \ .
\end{eqnarray}
We can directly compute the integral to find,
\begin{eqnarray}
    \mathcal{N}&=&\frac{2\sqrt{3}m}{\pi\sqrt{-R_-}}\Big[\Big(1-R_{+}\Big) K \Big(\frac{R_{+}}{R_{-}}\Big)-R_{-} E \Big(\frac{R_{+}}{R_{-}}\Big)+(R_{-} -1)(R_{+} -1) ~ \Pi \Big(R_{+}|\frac{R_{+}}{R_{-}}\Big)\Big]  \ , \nonumber \\  
\end{eqnarray}
where $K(k), E(k)$ and $\Pi(n,k)$ are standard complete elliptic integrals of first, second and third kind respectively. Instead of working with this, we can made the expression a little simpler by taking the partial derivative of $\mathcal{N}$ with respect to $m$,
\begin{eqnarray}
    \frac{\partial \mathcal{N}}{\partial m} & = & \frac{2\sqrt{3}}{\pi}\int_{0}^{\sqrt{R_+}}\frac{x^2dx}{\sqrt{(x^2-R_-)(R_+-x^2)}} \nonumber \\ && \hspace{4.2cm} -\frac{4(\mathcal{J}+2m)}{\sqrt{3}\pi m}\int_{0}^{\sqrt{R_+}}\frac{dx}{(1-x^2)\sqrt{(x^2-R_-)(R_+-x^2)}} \nonumber
\\ && = \frac{2\sqrt{-3R_-}}{\pi} \left[K\left(\frac{R_+}{R_-}\right)-E\left(\frac{R_+}{R_-}\right)\right] -\cfrac{4(\mathcal{J}+2m)}{\sqrt{-3R_-}\pi m} ~ \Big[ \Pi \left(R_+|\frac{R_+}{R_-}\right)\Big] .
\end{eqnarray}
In the short string limit, i.e. when both the energy and angular momentum of the string are small, we can expand the above expression in  $\mathcal{E}$ and $\mathcal{J}$ to get the oscillation number
\begin{equation}
    \mathcal{N} = \mathcal{A}(\mathcal{J}) + \mathcal{E}^2 \mathcal{B}(\mathcal{J}) + \mathcal{O}(\mathcal{E}^4) \label{N eq}
\end{equation}
Inverting (\ref{N eq}) we can write the energy-oscillation number relation as,
\begin{eqnarray}
	    \mathcal{E} & =& [\mathcal{B}(\mathcal{J})]^{-1/2}\sqrt{\mathcal{N} - \mathcal{A}(\mathcal{J})}+\mathcal{O}(\mathcal{N} - \mathcal{A}(\mathcal{J}))^{3/2} \label{en1}
\end{eqnarray}
where,\\
\begin{eqnarray}
\mathcal{A}(\mathcal{J}) &=& \left(4m-\frac{15 (\mu ^2+1)}{2 m}-\frac{159
	\mu ^2}{4 m^3}\right)-\bigg(6 \log m-\frac{153 (\mu ^2+ 1)}{4 m^2}-
	\frac{33435 \mu ^2}{64 m^4}\bigg)\mathcal{J} - \nonumber \\ &&  \bigg(\frac{51}{2 m}+\frac{213 (\mu ^2+1)}{m^3}  + \frac{81465 \mu ^2}{16
		m^5}\bigg) \mathcal{J}^2 +  \Big(\frac{267}{4 m^2}+\frac{81015 (\mu ^2+1)}{64 m^4}+\frac{1422975 \mu ^2}{32 m^6}\Big) \mathcal{J}^3 \nonumber \\ && +\mathcal{O}(\mathcal{J}^4) \ , \label{y}
\end{eqnarray}
and
\begin{eqnarray}
		\mathcal{B}(\mathcal{J}) &=& \left(\frac{15}{2m} +\frac{159 \left(\mu ^2+1\right)}{4 m^3}+ \frac{10995 \mu ^2}{16 m^5}\right)- \left(\frac{153}{4 m^2} + \frac{33435 \left(\mu ^2+1\right)}{64 m^4} + \frac{911955 \mu ^2}{64 m^6} \right) \mathcal{J} \nonumber \\&& +  \left(\frac{213}{m^3}+\frac{81465\left(\mu ^2+1\right)}{16 m^5}+\frac{10899675 \mu ^2}{56 m^7}\right)\mathcal{J}^2 - \Big(\frac{81015}{64 m^4}+\frac{1422975 \left(\mu ^2+1\right)}{32m^6} \nonumber \\ && \hspace{5.5cm}+\frac{283301175 \mu ^2}{128 m^8}\Big) \mathcal{J}^3+\mathcal{O}(\mathcal{J}^4) \ . \label{a} 
\end{eqnarray}
We will compare these relations (\ref{y}), (\ref{a}) along with the relation (\ref{en1}) with the corresponding relations of I-brane in the next section (\ref{sec3}).

\section{Pulsating strings in I-brane}\label{sec3}
The geometry of I-brane arises when $k_1$ number of NS5-branes lying along $(0,1,\cdots,5)$ intersect $k_2$ number of NS5-branes lying along $(0,1,6,\cdots,9)$ directions in (0,1)-plane. If the branes are coincident, then the type IIB supergravity solution is given by the following metric, three form NS-NS fields and dilaton \cite{Itzhaki:2005tu}, as,
\begin{eqnarray}
    && ds^2= -(dx^0)^2 + (dx^1)^2 + H_1(y) \sum_{\alpha=2}^5 (dy^{\alpha})^2 + 
H_2(z) \sum_{p=6}^9 (dz^p)^2 \ , \nonumber \\ H_{\alpha\beta\gamma} &=& -\epsilon_{\alpha\beta\gamma\delta} \partial^{\delta} H_1(y) \ , ~~ H_{mnp}=-\epsilon_{mnpq}\partial^qH_2(z) \ , ~~ e^{2\Phi} = H_1(y)H_2(z) \ ,
\end{eqnarray}
where  the harmonic functions are $H_1(y) = 1+\frac{k_1l_s^2}{y^2}$ and $H_2(z) = 1+\frac{k_2l_s^2}{z^2}$ with $y= \sqrt{\sum_{\alpha=2}^5 (y^{\alpha})^2}$ and  $z= \sqrt{\sum_{p=6}^9 (z^{p})^2}$.
In the near horizon limit ($\frac{k_1l_s^2}{y^2}>>1$ and $\frac{k_2l_s^2}{z^2}>>1$), the metric and the NS-NS two form fields can be written as,
\begin{eqnarray}
    && ds^2 = -(dx^0)^2 + (dx^1)^2 + k_1l_s^2 \frac{dr_1^2}{r_1^2} + k_1l_s^2d\Omega_1^2 + k_2l_s^2 \frac{dr_2^2}{r_2^2} + k_2l_s^2d\Omega_2^2 \ , \nonumber \\ && B_{\phi_1\psi_1} = 2k_1l_s^2\sin^2\theta_1 d\phi_1\wedge d\psi_1 \ , ~~~  B_{\phi_2\psi_2} = 2k_2l_s^2\sin^2\theta_2 d\phi_2 \wedge d\psi_2 \ ,
\end{eqnarray}
where the three spheres $d\Omega_1, d\Omega_2$ are the volume elements on the sphere along $(y^2,\cdots,y^5)$ and $(z^6,\cdots,z^9)$ directions respectively, and are given by,
$$d\Omega_1^2 = d\theta_1^2 + \sin^2\theta_1 d\phi_1^2 + \cos^2\theta_1 d\psi_1^2 \ , ~~~ d\Omega_2^2 = d\theta_2^2 + \sin^2\theta_2 d\phi_2^2 + \cos^2\theta_2 d\psi_2^2 \ . $$
To proceed further we choose $l_s=1, k_1=k_2=N$, rescale $x^0 \to \sqrt{N}t, x^1 \to \sqrt{N}x$ and make the following change of variables
\begin{eqnarray}
    \rho_1 = \ln \frac{r_1}{\sqrt{N}} \ , ~~ \rho_2 = \ln \frac{r_2}{\sqrt{N}} \ . 
\end{eqnarray}
The final form of the metric and the background NS-NS field are given by,
\begin{eqnarray} \label{metric1}
    && ds^2 = N(-dt^2 + dx^2 + d\rho_1^2 + d\theta_1^2 + \sin^2\theta_1 d\phi_1^2 + \cos^2\theta_1 d\psi_1^2 + d\rho_2^2 + d\theta_2^2 \nonumber \\ && + \sin^2\theta_2 d\phi_2^2 + \cos^2\theta_2 d\psi_2^2) \ , ~~ b_{\phi_1\psi_1} = 2N \sin^2\theta_1 \ , ~~ b_{\phi_2\psi_2} = 2N \sin^2\theta_2 \ .
\end{eqnarray}

\subsection{String equations of motion}
For studying a generic class of pulsating strings in I-brane background (\ref{metric1}) using the Polyakov action of string as before, we use the following ansatz,
\begin{eqnarray}
    && t = t(\tau) = \kappa\tau  \ , ~~~ x = \sigma \ , ~~~ \rho_1 = \mu_1\sigma \ , ~~~ \rho_2 = \mu_2\sigma \ , ~~~ \theta_1=\theta_1(\tau) \ , \nonumber \\ && \theta_2 = \theta_2(\tau)  \ , ~~~ \phi_1 = m_1\sigma \ , ~~~ \phi_2 = m_2\sigma \ , ~~~  \psi_1 = \psi_1(\tau) \ , ~~~ \psi_2 = \psi_2(\tau) \ , \label{ansatz} \nonumber \\
\end{eqnarray}
Solving $t$, $\psi_1$ and $\psi_2$ equations we obtain,
\begin{eqnarray}
    \dot{t} &=& \kappa \ , \nonumber \\ \dot{\psi}_1 &=& \frac{c_1 + 2m_1\sin^2\theta_1}{\cos^2\theta_1}  \ , \nonumber \\  \dot{\psi}_2 &=& \frac{c_2+ 2m_2\sin^2\theta_2}{\cos^2\theta_2} \ . 
\end{eqnarray}
The equations for $x$, $\rho_{1,2}$ and $\phi_{1,2}$ are satisfied trivially. The other non-trivial contributions come from $\theta_{1,2}$ equations, solving them we get,
\begin{eqnarray}
    \ddot{\theta}_1 &=& \Big[3m_1^2 - \frac{(c_1+2m_1)^2}{\cos^4\theta_1}\Big] \sin\theta_1\cos\theta_1 \ ,  \\ \ddot{\theta}_2 &=& \Big[3m_2^2 - \frac{(c_2+2m_2)^2}{\cos^4\theta_2}\Big] \sin\theta_2\cos\theta_2 \ .
\end{eqnarray}
Integrating the above equations, we get,
\begin{eqnarray}
    \dot{\theta}_1^2 &=& c_3 + 3m_1^2\sin^2\theta_1 - \frac{(c_1+2m_1)^2}{\cos^2\theta_1} \ , \label{th1} \\ \dot{\theta}_2^2 &=& c_4 + 3m_2^2\sin^2\theta_2 - \frac{(c_2+2m_2)^2}{\cos^2\theta_2} \ , \label{th2}
\end{eqnarray}
where $c_{1,2,3,4}$ are constants.
\medskip

Now, the Virasoro constraint $g_{MN}(\partial_{\tau}X^M\partial_{\tau}X^N + \partial_{\sigma}X^M\partial_{\sigma}X^N)=0$, gives 
\begin{eqnarray}
    \dot{\theta}_1^2 + \dot{\theta}_2^2 &=& \alpha^2 + 4m_1^2 + 4m_2^2 + 4c_1m_1 + 4c_2m_2  \nonumber \\ && + 3m_1^2\sin^2\theta_1 + 3m_2^2\sin^2\theta_2 - \frac{(c_1+2m_1)^2}{\cos^2\theta_1} - \frac{(c_2+2m_2)^2}{\cos^2\theta_2} \ , \label{Vir1}
\end{eqnarray}
where $\alpha^2 = \kappa^2 -1 - \mu_1^2 - \mu_2^2$. The other Virasoro $g_{MN}\partial_{\tau}X^M\partial_{\sigma}X^N=0$ is trivially satisfied. Adding (\ref{th1}) and (\ref{th2}) we get,
\begin{equation}
    \dot{\theta}_1^2 + \dot{\theta}_2^2 = c_3 + c_4+ 3m_1^2\sin^2\theta_1 + 3m_2^2\sin^2\theta_2 - \frac{(c_1+2m_1)^2}{\cos^2\theta_1} - \frac{(c_2+2m_2)^2}{\cos^2\theta_2} \ , \label{th3}
\end{equation} 
Comparing the Virasoro constraint (\ref{Vir1}) with the equation of motion (\ref{th3}), we get the following relation between various constants,
\begin{equation}
    c_3+c_4 = \alpha^2 + 4m_1^2 + 4m_2^2 + 4c_1m_1 + 4c_2m_2 \label{cond1} \ .
\end{equation}
\medskip

To determine the conserved charges associated to the string motion, we write down the full form of the sigma model action in background \eqref{metric1},
\begin{eqnarray}
   S &=& -\frac{\sqrt{\lambda}}{4\pi} \int d\tau d\sigma \Big[ \dot{t}^2 +1 + \mu_1^2 + \mu_2^2 - \dot{\theta}_1^2 - \dot{\theta}_2^2 + m_1^2\sin^2\theta_1 + m_2^2\sin^2\theta_2 \nonumber \\ && -\dot{\psi}_1^2\cos^2\theta_1 - \dot{\psi}_2^2\cos^2\theta_2 + 4m_1\dot{\psi}_1\sin^2\theta_1 + 4m_2\dot{\psi}_2\sin^2\theta_2 \Big] \ .
\end{eqnarray}
From this action we can easily determine the conserved energy ($E$) and angular momenta ($J_1$) and ($J_2$) of the two orthogonal spheres, as following,
\begin{eqnarray}
    E &=& - \int \frac{\partial\mathcal{L}}{\partial \dot{t}} ~d\sigma = \sqrt{\lambda}\kappa \ , \nonumber \\
J_{1} &=&  \int \frac{\partial\mathcal{L}}{\partial \dot{\psi}_1} ~d\sigma = \sqrt{\lambda} c_1 \ , \nonumber \\ J_{2} &=&  \int \frac{\partial\mathcal{L}}{\partial \dot{\psi}_2} ~d\sigma = \sqrt{\lambda} c_2  \ . 
\end{eqnarray}
As before, rescaling them as,
\begin{eqnarray}
    \mathcal{E} = \frac{E}{\sqrt{\lambda}} = \kappa \ , ~~~ \mathcal{J}_1 = \frac{J_1}{\sqrt{\lambda}} = c_1 \ , ~~~ \mathcal{J}_2 = \frac{J_2}{\sqrt{\lambda}} = c_2 \ .
\end{eqnarray}
Now expressing $\kappa$, $c_1$ and $c_2$ in terms of these conserved charges we can express the constraint equation (\ref{cond1}) as,
\begin{equation}
    c_3 + c_4 = \mathcal{E}^2 -1 - \mu_1^2 - \mu_2^2 + 4m_1^2 + 4m_2^2 + 4m_1\mathcal{J}_1+ 4	m_2\mathcal{J}_2 \ . \label{cond2}
\end{equation}
As we have only the condition (\ref{cond2}), this is not sufficient to separate $c_3$ and $c_4$ in terms of the conserved charges. But, to plot the string profile and to expand the oscillation numbers in terms of $\mathcal{E}$ we need to separate $c_3$ and $c_4$. In the following, we will separate them by hand using some arguments on the energy distributions among the spheres of I-brane.

From the form of the constraint (\ref{condb}) of NS5-brane and from the ansatz (\ref{ansatz}) of I-brane, we can easily guess $\mu_1, m_1,\mathcal{J}_1$ belongs to the sphere $d\Omega_1$ while $\mu_2, m_2,\mathcal{J}_2$ belongs to the other sphere $d\Omega_2$ of I-brane. First we will distribute the remaining quantity ($\mathcal{E}^2 - 1$) equally to both the spheres and we will call this situation as the energy is equally distributed among the spheres of I-brane. So, for equal energy distribution case we separate $c_3$ and $c_4$ as,
\begin{eqnarray}
    c_3 = \frac{\mathcal{E}^2 - 1}{2} - \mu_1^2 + 4m_1^2 + 4m_1\mathcal{J}_1 \ , ~~~ c_4 = \frac{\mathcal{E}^2 - 1}{2} - \mu_2^2 + 4m_2^2 + 4m_2\mathcal{J}_2 \ ,
\end{eqnarray}
such that the constraint (\ref{cond2}) is satisfied.

Next we will distribute energy unequally among the spheres of I-brane by considering a factor $f$ ($0<f<1$). In this case we will distribute $f$ of the total energy to the sphere $d\Omega_1$ and rest of it to the sphere $d\Omega_2$, hence, we can separate $c_3$ and $c_4$ respecting the constraint (\ref{cond2}) as,
\begin{eqnarray}
    c_3 = f(\mathcal{E}^2 - 1) - \mu_1^2 + 4m_1^2 + 4m_1\mathcal{J}_1 \ , ~~~ c_4 = (1-f)(\mathcal{E}^2 - 1)- \mu_2^2 + 4m_2^2 + 4m_2\mathcal{J}_2 \ . \nonumber \\
\end{eqnarray}

\subsection{String profile}
From (\ref{th1}) we find as $\theta_1$ varies from $0$ to $\pi/2$, $\dot{\theta}_1^2$ varies from $c_3 - (\mathcal{J}_1+2m_1)^2$ to infinity which looks like the equations of motion of a particle moving in an effective potential $V(\theta_1)$, where $\theta_1$ rotates between a minimal and a maximal value. Note that the $\theta_1$ equation can be written as,
\begin{equation}
    \ddot{x}_1 + 3m_1^2[-(R_{1-}+R_{1+})x_1+2x_1^3] = 0 \ , \label{x1}
\end{equation}
where $x_1=\sin\theta_1$.

Again from (\ref{th2}) we find that $\dot{\theta}_2^2$ varies from $c_4 - (\mathcal{J}_2+2m_2)^2$ to infinity as $\theta_2$ varies from $0$ to $\pi/2$. This again looks like the particle moving in another effective potential $V(\theta_2)$, and $\theta_2$ rotates between a minimal and a maximal value. Using $x_2=\sin\theta_2$, $\theta_2$ equation can be written as,
\begin{equation}
    \ddot{x}_2 + 3m_2^2[-(R_{2-}+R_{2+})x_2 + 2x_2^3] = 0 \ , \label{x2}
\end{equation}
Thus, if both $(R_{1-}+R_{1+})$ and $(R_{2-}+R_{2+})$ are negative quantities, then we get two independent Duffing oscillator equations without damping and deriving terms. Integrating (\ref{x1}) and (\ref{x2}) we get,
\begin{eqnarray}
    \dot{x}_1^2 &=& 3m_1^2(x_1^2 - R_{1-})(R_{1+}-x_1^2) \ , \label{th4} \\ \dot{x}_2^2 &=&  3m_2^2(x_2^2 - R_{2-})(R_{2+}-x_2^2) \ , \label{th5}
\end{eqnarray}
where 
\begin{eqnarray}
    R_{1\pm} &=& \frac{-(c_3 - 3m_1^2) \pm \sqrt{(c_3 + 3m_1^2)^2 - 12 m_1^2 (\mathcal{J}_1 + 2m_1)^2}}{6m_1^2} \ , \nonumber \\ R_{2\pm} &=& \frac{-(c_4 - 3m_2^2) \pm \sqrt{(c_4 + 3m_2^2)^2 - 12 m_2^2 (\mathcal{J}_2 + 2m_2)^2}}{6m_2^2} \ .
\end{eqnarray}
On further integrating (\ref{th4}) and (\ref{th5}), we can express the solutions in terms of Jacobi functions provided the initial conditions $x_1(0)=x_2(0)=0$:
\begin{eqnarray}
    \sin \theta_1(\tau) &=& \sqrt{\frac{-R_{1+}R_{1-}}{R_{1+} -R_{1-}}}sd\Big( \sqrt{3(R_{1+}-R_{1-})}m_1\tau ~~,~~\sqrt{\frac{R_{1+}}{R_{1+}-R_{1-}}} \Big) \, \\ 
      \sin \theta_2(\tau) &=&\sqrt{\frac{-R_{2+}R_{2-}}{R_{2+} -R_{2-}}}sd\Big(\sqrt{3(R_{2+}-R_{2-})}m_2\tau ~~,~~\sqrt{\frac{R_{2+}}{R_{2+}-R_{2-}}} \Big)
\end{eqnarray}
Thus in this case we obtain two copies of string profiles arising from the two spheres of I-brane. These profiles are independent of each other and can exist simultaneously. We also note that each profile is exactly equivalent to the profile (\ref{profile1}) obtained for the NS5-brane.

Again, using the property of Jacobi functions and taking only the real period, we find the condition for time-periodic solution for $\theta_1$ and $\theta_2$ is,
\begin{eqnarray}
	0 < \frac{R_{1+}}{R_{1+}-R_{1-}} <1  \ , ~~~  0 < \frac{R_{2+}}{R_{2+}-R_{2-}} < 1 .
\end{eqnarray}
This translates to the following inequality,
\begin{eqnarray}
		\mathcal{E}^2 - 1 - \mu_1^2 - \mu_2^2- \mathcal{J}_1^2 - \mathcal{J}_2^2 > 0 \ , \label{cond3}
\end{eqnarray}
which is the constraint on the conserved charges so that the string has pulsating motion. Hence, we obtained the pulsating string profile which can pulsate on both the spheres independently and simultaneously. The condition (\ref{cond3}) is also in tune with limits of $\dot{\theta}_1^2$ and $\dot{\theta}_2^2$ of (\ref{th1}) and (\ref{th2}).
\medskip

\begin{figure}[h!]
	\centering
	\subfigure{\includegraphics[width=0.25\linewidth]{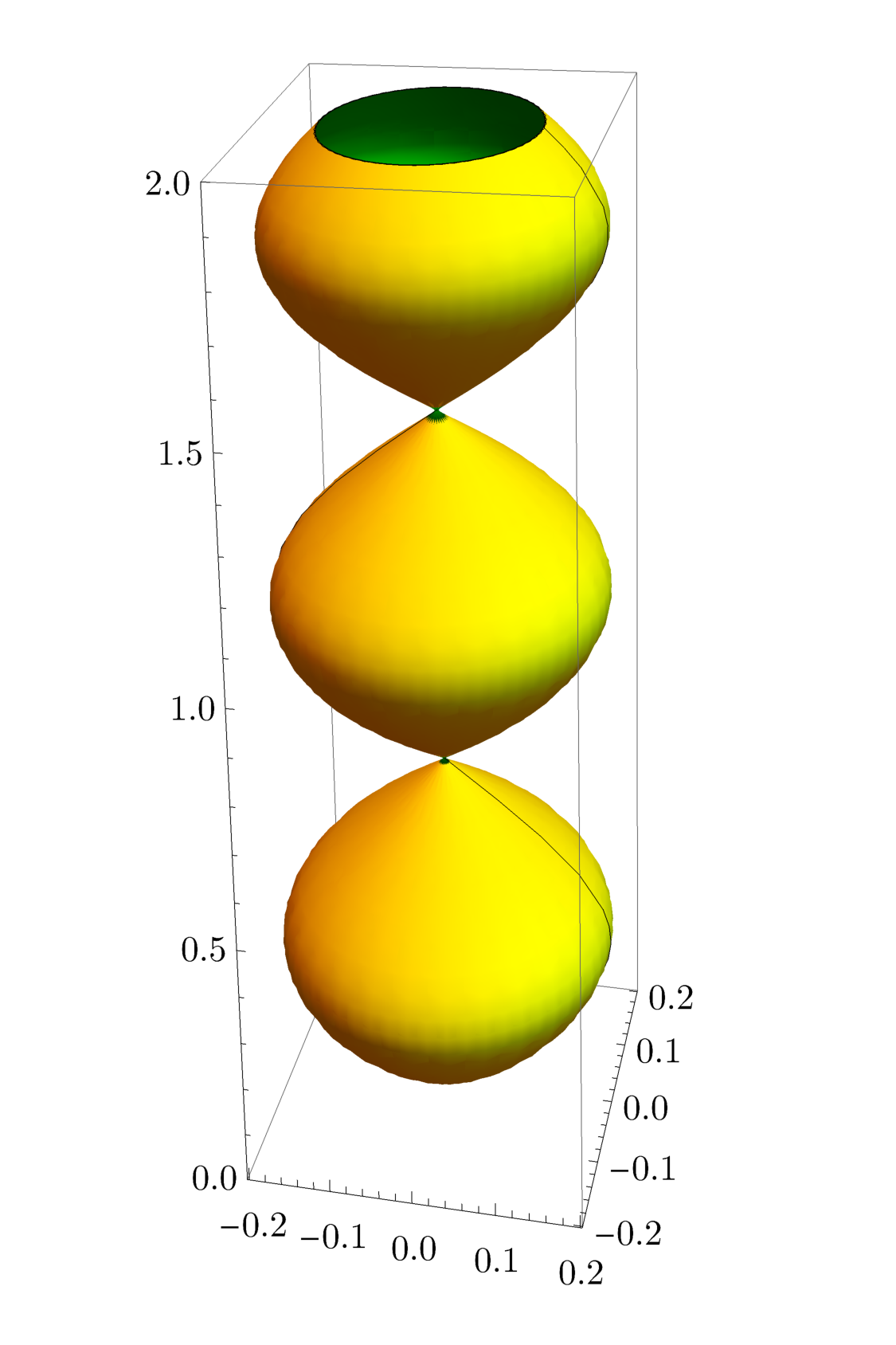}} 
	\subfigure{\includegraphics[width=0.25\linewidth]{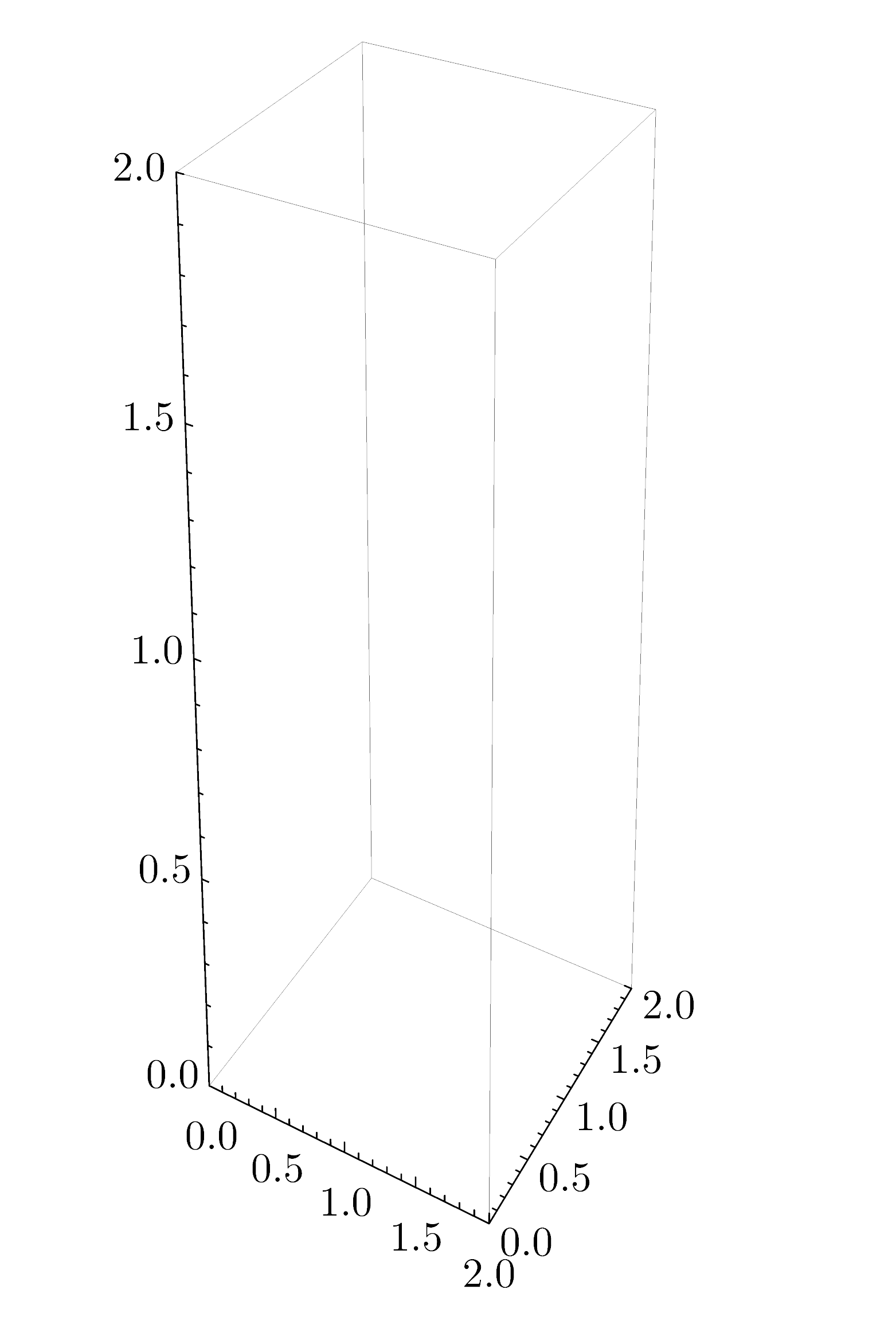}} 
	\caption{Pulsating string profile on I-brane for $f=1$ with $\mathcal{E}=2$, $m_1=m_2=3$,~$\mathcal{J}_1=\mathcal{J}_2=0.05$,~$\mu_1=\mu_2=1$ .} \label{fig 4}
\end{figure}
	\begin{figure}[h!]
	\centering
	\subfigure{\includegraphics[width=0.29\linewidth]{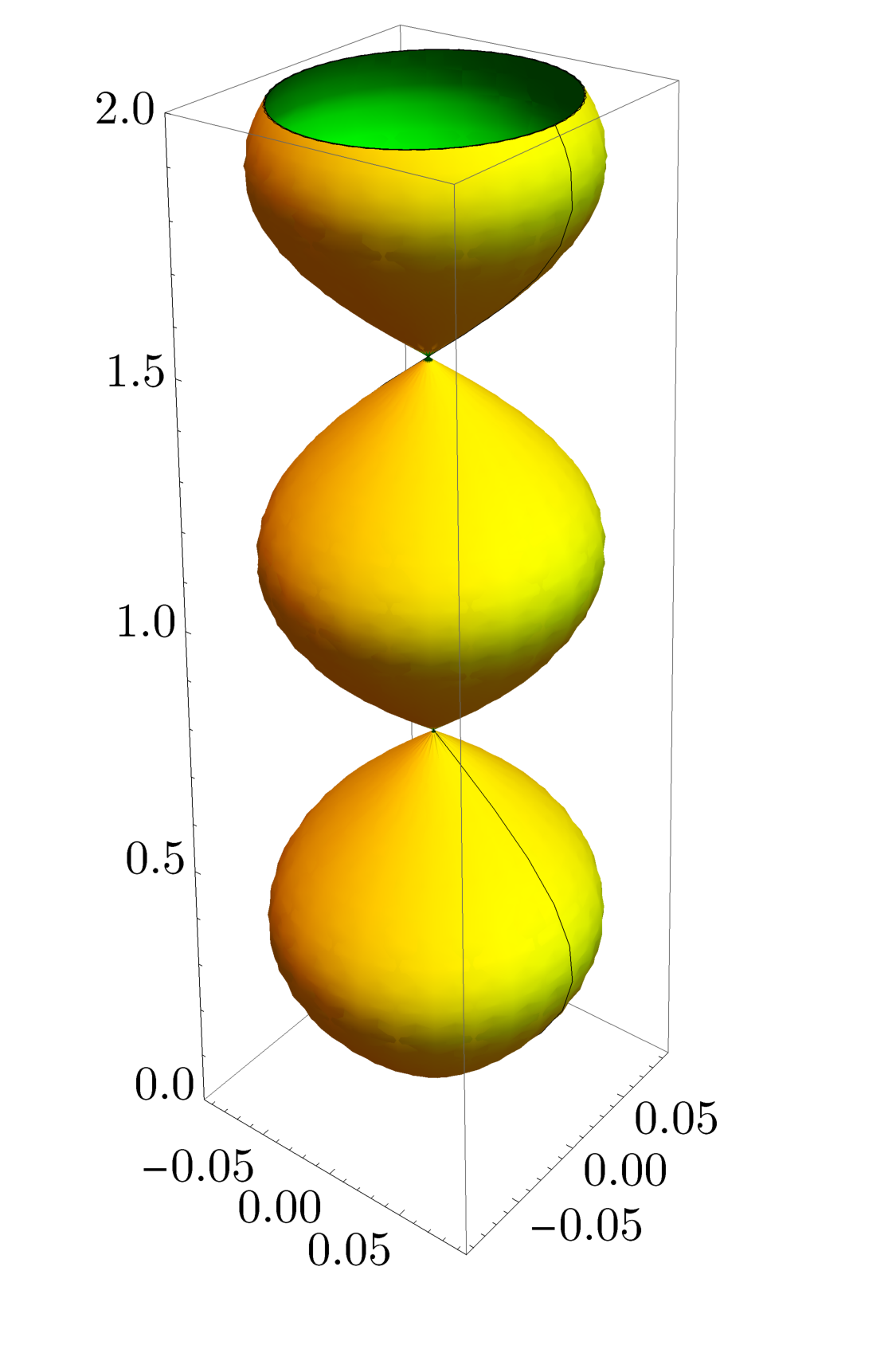}}
	\subfigure{\includegraphics[width=0.29\linewidth]{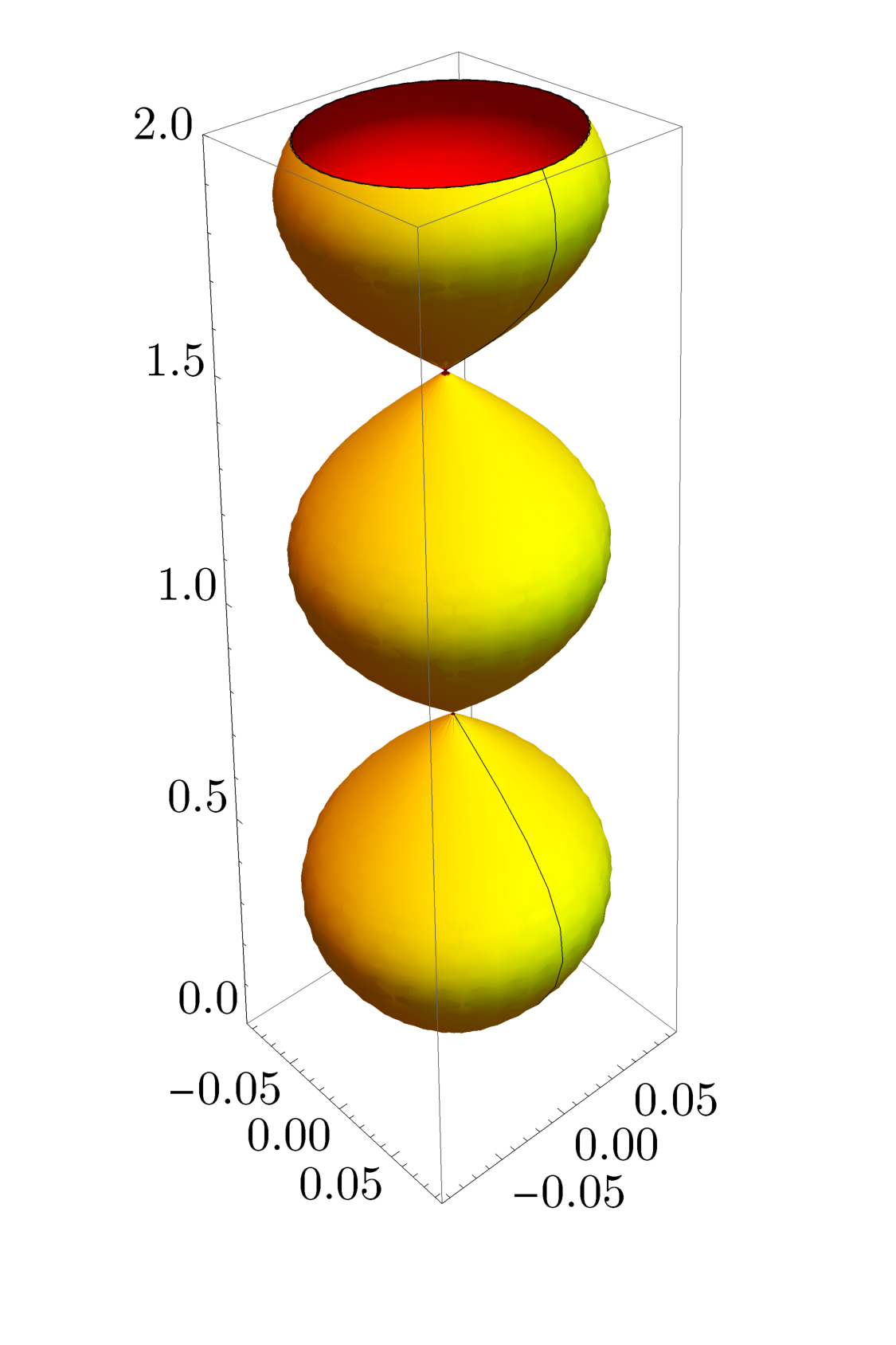}} 
	\caption{Pulsating string profile for $f=0.5$ with $\mathcal{E}=2$,  $m_1=m_2=3$,~$\mathcal{J}_1=\mathcal{J}_2=0.5$,~$\mu_1=\mu_2=1$ .} \label{fig 5}
\end{figure}
	\begin{figure}[h!]
	\centering
	\subfigure{\includegraphics[width=0.25\linewidth]{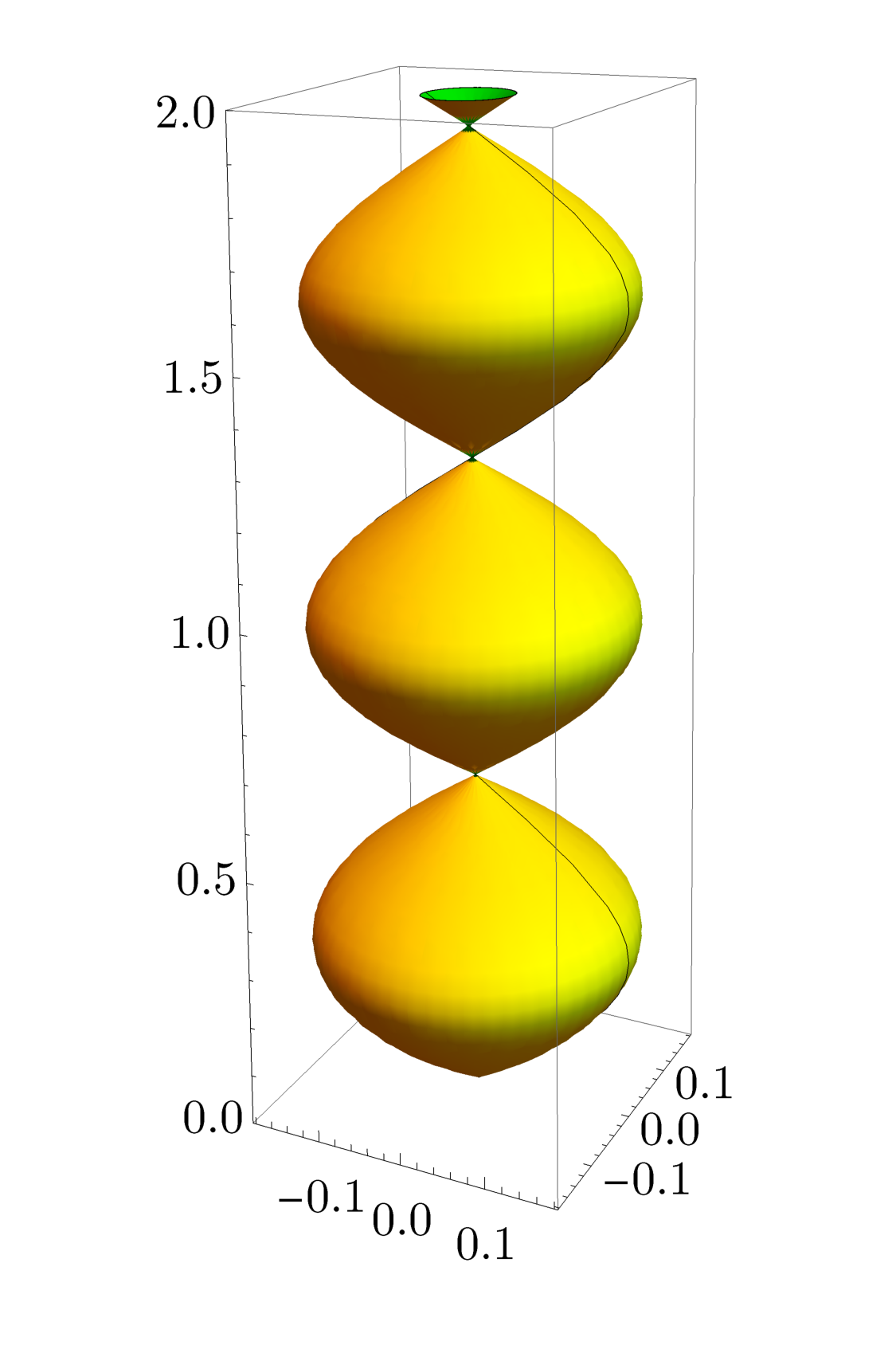}}
	\subfigure{\includegraphics[width=0.25\linewidth]{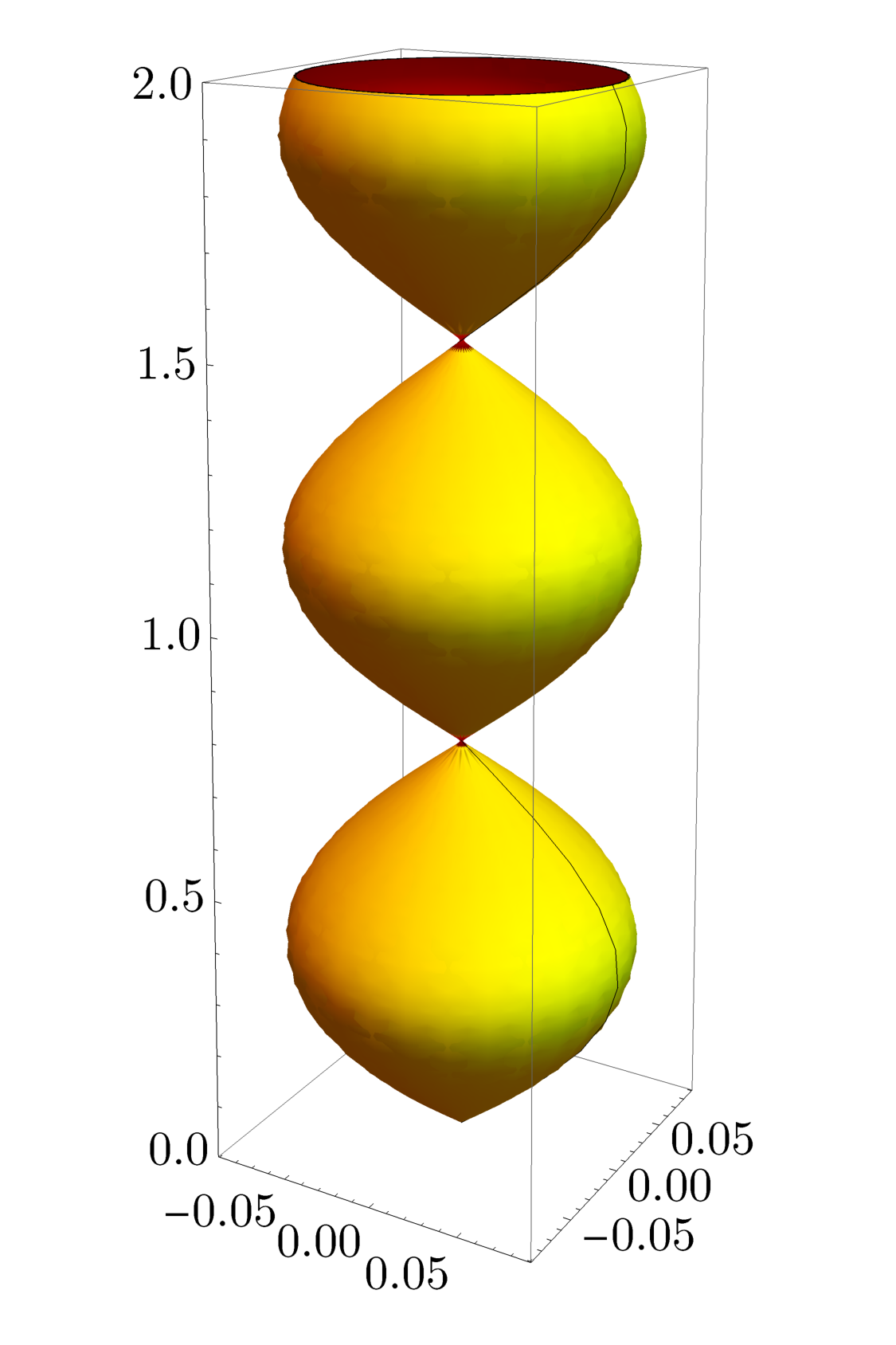}} 
	\caption{Pulsating string profile for $f=0.7$ with $\mathcal{E}=2.5$,  $m_1=m_2=3$,~$\mathcal{J}_1=\mathcal{J}_2=0.5$,~$\mu_1=\mu_2=1$ .} \label{fig 6}
\end{figure}
Figure (\ref{fig 4}) shows the string profile of I-brane background for $f=1$, in this case the energy is fully distributed to the sphere $d\Omega_1$ and there is no energy available to the other sphere $d\Omega_2$. In this case, the string have the pulsating motion only on the sphere $d\Omega_1$ and as there is no energy available to the other sphere $d\Omega_2$, string don't have any pulsating motion there although we have used non-zero values of angular momenta and winding numbers. Instead, if we had used $f=0$ then we would see the string to pulsate only on the sphere $d\Omega_2$. Thus, if we switch off the energy in one of the sphere of I-brane, then the string profile behave similar to that of NS5-brane on the other sphere. In figure (\ref{fig 5}) we presented the string profile for $f=0.5$, in this case energy is equally distributed among the spheres and we see the string have pulsating motion on both the spheres simultaneously. We can also note for equal energy distribution case as the winding numbers $m_1$ and $m_2$ are equal, the string is pulsating simultaneously on both the sphere with same amplitude. Instead, if we had used $m_1 \ne m_2$, then we would expect string to pulsate with different amplitude by forming unequal number of lobes as we saw in case of NS5-branes. Figure (\ref{fig 6}) shows string profile for $f=0.7$, here $70$ percent of energy is distributed to the sphere $d\Omega_1$ and $30$ percent to the sphere $d\Omega_2$. We see with more energy the string on the sphere $d\Omega_1$ pulsate with larger amplitude, simultaneously the string is pulsating with smaller amplitude on the other sphere $d\Omega_2$ as less energy is available there. We can also see that first sphere have larger number of lobes compared to the second sphere for same extent of worldsheet time instead of same value of winding numbers $m_1=m_2$.
\medskip
 
We can also find the dynamics of the string along the $\psi_1$ and $\psi_2$ direction by integrating $\frac{d\psi_1}{d\theta_1}$ and $\frac{d\psi_2}{d\theta_2}$ which have the form,
\begin{eqnarray}
    \frac{d\psi_1}{d\theta_1} &=& \frac{\mathcal{J}_1 + 2m_1\sin^2\theta_1}{\sqrt{3}m_1\cos\theta_1 \sqrt{(\sin^2\theta_1 - R_{1-})(R_{1+}-\sin^2\theta_1 )}} \ , \nonumber \\ \frac{d\psi_2}{d\theta_2} &=& \frac{\mathcal{J}_2 + 2m_2\sin^2\theta_2}{\sqrt{3}m_2\cos\theta_2 \sqrt{(\sin^2\theta_2 - R_{2-})(R_{2+}-\sin^2\theta_2 )}} \ .
\end{eqnarray}
These can be integrated to find $\psi_1$ and $\psi_2$ in terms of standard elliptic integrals,
\begin{eqnarray}
    \psi_1(\tau) &=& \frac{-1}{\sqrt{-3R_{1-}}} \Big[\frac{\mathcal{J}_1 + 2m_1}{m_1} \Pi \Big(R_{1+}, \arcsin\Big(\frac{\sin\theta_1(\tau)}{\sqrt{R_{1+}}}\Big), \frac{R_{1+}}{R_{1-}}\Big) \nonumber \\ && + 2F \Big( \arcsin\Big(\frac{\sin\theta_1(\tau)}{\sqrt{R_{1+}}}\Big), \frac{R_{1+}}{R_{1-}} \Big) \Big] \ , \\ \psi_2(\tau) &=& \frac{-1}{\sqrt{-3R_{2-}}} \Big[\frac{\mathcal{J}_2 + 2m_2}{m_2} \Pi \Big(R_{2+}, \arcsin\Big(\frac{\sin\theta_2(\tau)}{\sqrt{R_{2+}}}\Big), \frac{R_{2+}}{R_{2-}}\Big) \nonumber \\ && + 2F \Big( \arcsin\Big(\frac{\sin\theta_2(\tau)}{\sqrt{R_{2+}}}\Big), \frac{R_{2+}}{R_{2-}} \Big) \Big] \ .
\end{eqnarray}

\subsection{Oscillation Number}
As we did in the previous section here also we will use Bohr-Sommerfeld like quantization procedure and write down the oscillation numbers using the canonical momenta conjugate to $\theta_1$ and $\theta_2$ as follows,
\begin{eqnarray}
    N_1 &=& \sqrt{\lambda}\mathcal{N}_1 = \frac{\sqrt{\lambda}}{2\pi} \oint d\theta_1 \Pi_{\theta_1} \nonumber \\ && = \frac{\sqrt{\lambda}}{2\pi} \oint d\theta_1 \sqrt{c_3 + 3m_1^2\sin^2\theta_1 - \frac{(\mathcal{J}_1 + 2m_1)^2}{\cos^2\theta_1}} \ , \\ N_2 &=& \sqrt{\lambda}\mathcal{N}_2 = \frac{\sqrt{\lambda}}{2\pi} \oint d\theta_2 \Pi_{\theta_2} \nonumber \\ && = \frac{\sqrt{\lambda}}{2\pi} \oint d\theta_2 \sqrt{c_4 + 3m_2^2\sin^2\theta_2 - \frac{(\mathcal{J}_2 + 2m_2)^2}{\cos^2\theta_2}} \ .
\end{eqnarray}
Again, taking $\sin\theta_1=x_1$ and $\sin\theta_2=x_2$ and choosing the proper limits and the above integrals transform to,
\begin{eqnarray}
    \mathcal{N}_1 &=& \frac{2\sqrt{3}m_1}{\pi} \int_0^{\sqrt{R_{1+}}} \frac{\sqrt{(x_1^2 - R_{1-})( R_{1+}-x_1^2 )}}{1-x_1^2} dx_1 \ , \\ \mathcal{N}_2 &=& \frac{2\sqrt{3}m_2}{\pi} \int_0^{\sqrt{R_{2+}}} \frac{\sqrt{(x_2^2 - R_{2-})( R_{2+}-x_2^2 )}}{1-x_2^2} dx_2 \ .
\end{eqnarray}
Directly computing the integrals, we find,
\begin{eqnarray}
    \mathcal{N}_1 &=& \frac{2\sqrt{3}m_1}{\pi\sqrt{-R_{1-}}} \Big[\Big(1-R_{1+}\Big) K \Big(\frac{R_{1+}}{R_{1-}}\Big)-R_{1-} E \Big(\frac{R_{1+}}{R_{1-}}\Big)+(R_{1-} -1)(R_{1+} -1) ~ \Pi \Big(R_{1+}|\frac{R_{1+}}{R_{1-}}\Big)\Big]  \ , \nonumber \\  \mathcal{N}_2 &=& \frac{2\sqrt{3}m_2}{\pi\sqrt{-R_{2-}}} \Big[\Big(1-R_{2+}\Big) K \Big(\frac{R_{2+}}{R_{2-}}\Big)-R_{2-} E \Big(\frac{R_{2+}}{R_{2-}}\Big)+(R_{2-} -1)(R_{2+} -1) ~ \Pi \Big(R_{2+}|\frac{R_{2+}}{R_{2-}}\Big) \Big] \nonumber  .
\end{eqnarray}
As before, in order to make the expressions little simpler we take the partial derivative of $\mathcal{N}_1$ with respect to $m_1$,
\begin{eqnarray}
   && \frac{\partial \mathcal{N}_1}{\partial m_1} = \cfrac{2\sqrt{3}}{\pi}\int_{0}^{\sqrt{R_{1+}}}\cfrac{x_1^2dx_1}{\sqrt{(x_1^2-R_{1-})(R_{1+}-x_1^2)}} \nonumber \\ && \hspace{4.2cm}-\frac{4(\mathcal{J}_1+2m_1)}{\sqrt{3}\pi m_1}\int_{0}^{\sqrt{R_{1+}}}\cfrac{dx_1}{(1-x_1^2)\sqrt{(x_1^2-R_{1-})(R_{1+}-x_1^2)}} \nonumber \vspace{1cm}\\ &&  = \cfrac{2\sqrt{-3R_{1-}}}{\pi} \left[K\left(\frac{R_{1+}}{R_{1-}}\right)-E\left(\frac{R_{1+}}{R_{1-}}\right)\right] -\cfrac{4(\mathcal{J}_1+2m_1)}{\sqrt{-3R_{1-}}\pi m_1}~\Big[ \Pi \left(R_{1+}|\frac{R_{1+}}{R_{1-}}\right)\Big] ~~.
\end{eqnarray}
Again, taking the partial derivative of $\mathcal{N}_2$ with respect to $m_2$ we get,
\begin{eqnarray}
&&	\frac{\partial \mathcal{N}_2}{\partial m_2} = \cfrac{2\sqrt{3}}{\pi}\int_{0}^{\sqrt{R_{2+}}}\cfrac{x_2^2dx_2}{\sqrt{(x_2^2-R_{2-})(R_{2+}-x_2^2)}}\nonumber \\ && \hspace{4.2cm}-\frac{4(\mathcal{J}_2 + 2m_2)}{\sqrt{3}\pi m_2}\int_{0}^{\sqrt{R_{2+}}}\cfrac{dx_2}{(1-x_2^2)\sqrt{(x_2^2-R_{2-})(R_{2+}-x_2^2)}} \nonumber \vspace{1cm}\\ && = \cfrac{2\sqrt{-3R_{2-}}}{\pi} \left[K\left(\frac{R_{2+}}{R_{2-}}\right)-E\left(\frac{R_{2+}}{R_{2-}}\right)\right] -\cfrac{4(\mathcal{J}_2+2m_2)}{\sqrt{-3R_{2-}}\pi m_2}~ \Big[ \Pi \left(R_{2+}|\frac{R_{2+}}{R_{2-}}\right)\Big]~~.
\end{eqnarray}
In the following we will discuss the expansion of the above expressions under short string limit for different kinds of energy distribution situations.
\subsubsection{Equal energy distribution case}
Assuming the energy is equally distributed on both the spheres. In the short string limit, i.e. when both the energy and angular momentum of the string are small, we can expand the above expressions as,
\begin{eqnarray}
    \mathcal{N}_1 &=& \mathcal{A}_1(\mathcal{J}_1) + \mathcal{E}^2 \mathcal{B}_1(\mathcal{J}_1) + \mathcal{O}(\mathcal{E}^4) \ , \label{N1 eq} \\ \mathcal{N}_2 &=& \mathcal{A}_2(\mathcal{J}_2) + \mathcal{E}^2 \mathcal{B}_2(\mathcal{J}_2) + \mathcal{O}(\mathcal{E}^4) \ . \label{N2 eq} 
\end{eqnarray}
Adding (\ref{N1 eq}) and (\ref{N2 eq}) and inverting the added series to get the expression for energy in terms of the conserved quantities as,
\begin{eqnarray}
	\mathcal{E} &=& [\mathcal{B}_1(\mathcal{J}_1) + \mathcal{B}_2(\mathcal{J}_2)]^{-1/2}\sqrt{\mathcal{N}_1 + \mathcal{N}_2 - \mathcal{A}_1(\mathcal{J}_1) - \mathcal{A}_2(\mathcal{J}_2)} \nonumber \\ &&+ \mathcal{O}(\mathcal{N}_1 + \mathcal{N}_2 - \mathcal{A}_1(\mathcal{J}_1) - \mathcal{A}_2(\mathcal{J}_2))^{3/2} \label{en2}
\end{eqnarray}
where
\begin{eqnarray}
	\mathcal{A}_1(\mathcal{J}_1) &=& \left(4m_1-\frac{15}{4 m_1} - \frac{15 \mu _1^2}{2 m_1} - \frac{159 \mu _1^2}{8 m_1^3}\right) - \left(6 \log m_1-\frac{153}{8 m_1^2}-\frac{153 \mu _1^2}{4 m_1^2}-\frac{33435 \mu_1^2}{128 m_1^4}\right)\mathcal{J}_1 \nonumber \\ && - \left(\frac{51}{2 m_1}+\frac{213}{2 m_1^3}+\frac{213 \mu _1^2}{m_1^3}+\frac{81465 \mu _1^2}{32 m_1^5}\right)\mathcal{J}_1^2 + \bigg(\frac{267 }{4	m_1^2} + \frac{81015 }{128 m_1^4}+\frac{81015  \mu _1^2}{64 m_1^4}+\nonumber \\ && ~~~\frac{1422975  \mu _1^2}{64 m_1^6} \bigg)\mathcal{J}_1^3 +\mathcal{O}(\mathcal{J}_1^4) \ ,
		\label{Y1}
\end{eqnarray}
\begin{eqnarray}
	\mathcal{A}_2(\mathcal{J}_2) &=&  \left(4m_2-\frac{15}{4 m_2} - \frac{15 \mu _2^2}{2 m_2} - \frac{159 \mu _2^2}{8 m_2^3}\right) - \left(6 \log m_2-\frac{153}{8 m_2^2}-\frac{153 \mu _2^2}{4 m_2^2}-\frac{33435 \mu_2^2}{128 m_2^4}\right)\mathcal{J}_2 \nonumber \\ && - \left(\frac{51}{2 m_2}+\frac{213}{2 m_2^3}+\frac{213 \mu _2^2}{m_2^3}+\frac{81465 \mu _2^2}{32 m_2^5}\right)\mathcal{J}_2^2 + \bigg(\frac{267 }{4	m_2^2} + \frac{81015 }{128 m_2^4}+\frac{81015  \mu _2^2}{64 m_2^4}+\nonumber \\ && ~~~\frac{1422975  \mu _2^2}{64 m_2^6} \bigg)\mathcal{J}_2^3 +\mathcal{O}(\mathcal{J}_2^4) \ ,
		\label{Y2}
\end{eqnarray}
\begin{eqnarray}
	\mathcal{B}_1(\mathcal{J}_1)&=&\left(\frac{15}{4 m_1}+\frac{159}{16 m_1^3}+\frac{159 \mu _1^2}{8 m_1^3}+\frac{10995 \mu _1^2}{64 m_1^5}\right)-\bigg(\frac{153}{8m_1^2} + \frac{33435}{256 m_1^4} + \frac{33435 \mu _1^2}{128 m_1^4} +  \nonumber \\ &&\frac{911955 \mu _1^2}{256m_1^6}\bigg)\mathcal{J}_1+\bigg(\frac{213}{2m_1^3}+\frac{81465}{64 m_1^5}+\frac{81465 \mu _1^2}{32 m_1^5}+\cfrac{10899675 \mu_1^2}{224m_1^7}\bigg)\mathcal{J}_1^2 -\bigg(\frac{81015}{128m_1^4} \nonumber \\ &&+\cfrac{1422975}{128 m_1^6}+\cfrac{1422975 \mu _1^2}{64 m_1^6}
	+\cfrac{283301175 \mu _1^2}{512 m_1^8} \bigg)\mathcal{J}_1^3 +\mathcal{O}(\mathcal{J}_1^4) \ ,
		\label{A1J1}
\end{eqnarray}
\begin{eqnarray}
	\mathcal{B}_2(\mathcal{J}_2)&=& \left(\frac{15}{4 m_2}+\frac{159}{16 m_2^3}+\frac{159 \mu _2^2}{8 m_2^3}+\frac{10995 \mu _2^2}{64 m_2^5}\right)-\bigg(\frac{153}{8m_2^2} + \frac{33435}{256 m_2^4} + \frac{33435 \mu _2^2}{128 m_2^4} + \nonumber \\ && \frac{911955 \mu _2^2}{256m_2^6}\bigg)\mathcal{J}_2 +\bigg(\frac{213}{2m_2^3}+\cfrac{81465}{64 m_2^5}+\frac{81465 \mu _2^2}{32 m_2^5}+\cfrac{10899675 \mu_2^2}{224m_2^7}\bigg)\mathcal{J}_2^2 -\bigg(\cfrac{81015}{128m_2^4}\nonumber \\&&+\frac{1422975}{128 m_2^6}+\cfrac{1422975 \mu _2^2}{64 m_2^6}+\cfrac{283301175 \mu _2^2}{512 m_2^8} \bigg)\mathcal{J}_2^3 +\mathcal{O}(\mathcal{J}_2^4) \ .
		\label{A2J2}	
\end{eqnarray}
	
\subsubsection{Unequal energy distribution case}
In this case expanding oscillation numbers $\mathcal{N}_1$ and $\mathcal{N}_2$ when the corresponding energy and angular momentum are small and inverting the added series, we obtain the similar kind of relation as,
\begin{eqnarray}
	\mathcal{E} &=& [\mathcal{B}_1(\mathcal{J}_1) + \mathcal{B}_2(\mathcal{J}_2)]^{-1/2}\sqrt{\mathcal{N}_1 + \mathcal{N}_2 - \mathcal{A}_1(\mathcal{J}_1) - \mathcal{A}_2(\mathcal{J}_2)} \nonumber \\ &&+ \mathcal{O}(\mathcal{N}_1 + \mathcal{N}_2 - \mathcal{A}_1(\mathcal{J}_1) - \mathcal{A}_2(\mathcal{J}_2))^{3/2} \label{en3}
\end{eqnarray}
but now,
\begin{eqnarray}
		\mathcal{A}_1(\mathcal{J}_1) &=&\left(4 m_1-\frac{15 f}{2 m_1}-\frac{15 \mu_1^2}{2 m_1}-\frac{159 f \mu _1^2}{4	m_1^3}\right)- \bigg(6 \log
		m_1-\frac{153 f}{4 m_1^2}-\frac{153 \mu _1^2}{4 m_1^2}-\frac{33435 f \mu
		_1^2}{64 m_1^4}\bigg)\mathcal{J}_1\nonumber \\&& - \bigg(\frac{51}{2 m_1}+ \frac{213
		f}{m_1^3}+\frac{213 \mu _1^2}{m_1^3}+\frac{81465 f \mu _1^2}{16 m_1^5}\bigg)\mathcal{J}_1^2+ \Big(\frac{267}{4 m_1^2}+\frac{81015 f}{64 m_1^4}+\frac{81015 \mu _1^2}{64
		m_1^4} + \nonumber \\&& ~~~\frac{1422975 f \mu _1^2}{32 m_1^6}\Big)\mathcal{J}_1^3+\mathcal{O}(\mathcal{J}_1)^4 \ , \label{y1}
\end{eqnarray}
\begin{eqnarray}
	\mathcal{A}_2(\mathcal{J}_2) &=& \left(4 m_2-\frac{15 (1-f)}{2 m_2}-\frac{15 \mu _2^2}{2 m_2}-\frac{159 (1-f) \mu _2^2}{4 m_2^3}\right)-\bigg(6 \log m_2-\frac{153 (1-f)}{4m_2^2} -\frac{153 \mu _2^2}{4 m_2^2}\nonumber \\&& -\frac{33435 (1-f) \mu _2^2}{64 m_2^4}\bigg)\mathcal{J}_2- \left(\frac{51}{2 m_2}+\frac{213 \mu _2^2}{m_2^3}+\frac{213(1-f)}{m_2^3}+\frac{81465 (1-f) \mu _2^2}{16 m_2^5}\right)\mathcal{J}_2^2 +\nonumber \\&& \bigg(\frac{267}{4m_2^2}+\frac{81015 (1-f)}{64 m_2^4}+\frac{81015 \mu _2^2}{64 m_2^4}+\frac{1422975 (1-f) \mu_2^2}{32 m_2^6}\bigg)\mathcal{J}_2^3 +\mathcal{O}(\mathcal{J}_2)^4 \ , \label{y2}
\end{eqnarray}
\begin{eqnarray}
	\mathcal{B}_1(\mathcal{J}_1)&=&\left(\frac{15 f}{2 m_1}+\frac{159 f^2}{4 m_1^3}+\frac{159 f \mu
		_1^2}{4 m_1^3}+\frac{10995 f^2 \mu
		_1^2}{16 m_1^5}\right)- \bigg(\frac{153 f}{4 m_1^2}+\frac{33435 f^2}{64
		m_1^4}+ \frac{33435 f \mu _1^2}{64 m_1^4} +\nonumber \\&& \cfrac{911955 f^2 \mu _1^2}{64 m_1^6}\bigg)\mathcal{J}_1+ \left(\frac{213
		f}{m_1^3}+\frac{81465 f^2}{16 m_1^5}+\frac{81465 f \mu _1^2}{16 m_1^5}+\cfrac{10899675 f^2 \mu _1^2}{56
		m_1^7}\right)\mathcal{J}_1^2 - \nonumber \\ &&  \bigg(\frac{81015 f}{64 m_1^4}+\frac{1422975 f^2}{32 m_1^6}+ \cfrac{1422975 f
		\mu _1^2}{32 m_1^6}+\cfrac{283301175 f^2 \mu _1^2}{128 m_1^8}\bigg)\mathcal{J}_1^3+\mathcal{O}(\mathcal{J}_1)^4 \ , \label{a1j1}
\end{eqnarray}
\begin{eqnarray}
	\mathcal{B}_2(\mathcal{J}_2)&=&\left(\frac{15 (1-f)}{2 m_2}+\frac{159 (1-f)^2}{4 m_2^3}+\frac{159 (1-f) \mu _2^2}{4 m_2^3}+\frac{10995 (1-f)^2 \mu _2^2}{16m_2^5}\right)-\nonumber \\ && \bigg(\frac{153 (1-f)}{4 m_2^2}+\frac{33435 (1-f)^2}{64 m_2^4}+\frac{33435 (1-f) \mu_2^2}{64 m_2^4}+\frac{911955 (1-f)^2 \mu _2^2}{64 m_2^6}\bigg)\mathcal{J}_2+\nonumber \\ && \bigg(\frac{213(1-f)}{m_2^3}+\frac{81465 (1-f)^2}{16 m_2^5}+\frac{81465 (1-f)\mu _2^2}{16 m_2^5}+\frac{10899675 (1-f)^2 \mu _2^2}{56 m_2^7}\bigg)\mathcal{J}_2^2 \nonumber \\ && -\bigg(\frac{81015(1-f)}{64 m_2^4}+\frac{1422975 (1-f)^2}{32 m_2^6}+\frac{1422975 (1-f) \mu _2^2}{32m_2^6} \nonumber \\ && \hspace{5.5cm}+\cfrac{283301175 (1-f)^2 \mu _2^2}{128	m_2^8}\bigg)\mathcal{J}_2^3 +\mathcal{O}(\mathcal{J}_2)^4 \ .
	 \label{a2j2}
\end{eqnarray}

Now we will discuss how the most general relations (\ref{y1})-(\ref{a2j2}) along with (\ref{en3}) reduces to different cases for different values of $f$.
First we consider $f=1$ that is when energy is fully distributed to the sphere $d\Omega_1$ and no energy is there to excite the string in the other sphere $d\Omega_2$ as demonstrated in figure (\ref{fig 4}). In this case as the string don't have any kind of motion in the sphere $d\Omega_2$, it won't be unlogical if we consider $\mathcal{J}_2=0$, in addition if we choose $\mu_2=\sqrt{\frac{8}{15}}m_2$ then we can easily check from (\ref{y2}) and (\ref{a2j2}) we obtain $\mathcal{A}_2(0)=0$ and $\mathcal{B}_2(0)=0$. These will in turn imply $\mathcal{N}_2=0$. Also, the relations (\ref{y1}) and (\ref{a1j1}) will reduces to (\ref{y}) and (\ref{a}) respectively. Hence, in this case the energy-oscillation number relation (\ref{en3}) exactly reduces to the energy-oscillation number relation of NS5-brane (\ref{en1}). 

One can obtain, similar kind of reduction for $f=0$ when the energy is fully distributed to the second sphere and no energy is available to the first sphere. In this case string will not have any kind of motion in the first sphere and we can obtain $\mathcal{A}_1(0)=\mathcal{B}_1(0)=\mathcal{N}_1=0$ and (\ref{y2}) and (\ref{a2j2}) reduces to (\ref{y}) and (\ref{a}) respectively. Thus for $f=1$ or $f=0$, i.e., when we switch off the energy in one of the sphere of I-brane then not only the string profile but also the energy-oscillation number relation of I-brane reduces to that of NS5-brane.

When the energy is equally distributed among the spheres that is for $f=0.5$ as demonstrated in figure (\ref{fig 5}), one can easily verify that the expressions (\ref{y1}), (\ref{y2}), (\ref{a1j1}) and (\ref{a2j2}) reduces to (\ref{Y1}), (\ref{Y2}), (\ref{A1J1}) and (\ref{A2J2}) respectively. Hence, we can conclude the relations of  the most general case of unequally energy distribution given by the relations (\ref{y1}) - (\ref{a2j2}) are consistent with the corresponding relations of equally energy distribution.

\section{Summary and conclusion}\label{sec4}
In this paper, first we have obtained the pulsating string solution i.e., oscillating behaviour under some constraint among various conserved charges and windings numbers in the NS5-brane background. We have shown the string profile is only sensitive to the energy and the winding number $m$. We have found that for a given energy, number of lobes increases with increasing winding number $m$, but the amplitude of pulsating string decreases. This can be thought of as the worldsheet of the string having some fixed value of surface area or tension for a given energy. If we want to form many lobes from this surface area then definitely each lobes will be of smaller size. Again, we found with more energy string will oscillate with higher amplitude. This can be explained by assuming that with increasing energy the surface area of the worldsheet increases and with higher value of surface area they can form larger lobes.We have also seen for the same value of winding number $m$, number of lobes increases with the energy. This might be because of the fact that with increasing energy not only the surface area of the worldsheet but also the tension of worldsheet increases. This tension will try to squeeze the lobes along the worldsheet time and hence form larger number of lobes within certain value of worldsheet time. For comparing we also determine the oscillation number and expand it under short string limit and obtain the energy oscillation number relation for pulsating string in NS5-brane background.
\medskip

Next, we have obtained pulsating string solution in I-brane background. Here we have found two copies of similar kind string profiles arising from the two spheres and not surprisingly each profile is equivalent to that of string profile in NS5-brane background. We have argued that in the near horizon geometry of I-brane background, string can have pulsating motion simultaneously and independently on both the spheres. In order to plot string profile and expand oscillation number we distribute energy equally and unequally among the spheres.  When energy is equally distributed among the spheres the string was shown to be pulsating on both the spheres simultaneously with same amplitude for $m_1=m_2$. When energy is unequally distributed, the string was shown to pulsate with larger amplitude on the sphere bearing higher energy and simultaneously with smaller amplitude on the sphere bearing lower energy. We have also shown when we cut off the energy from one of the spheres of I-brane then the string pulsates only on the other sphere. Finally , We have also expanded the oscillation number under short string limit and obtain the energy-oscillation number relation. 
\medskip

For complete understanding of these string solutions, one would look at  the possible dual gauge theory of these string theory backgrounds.  Recently, the authors of \cite{Nunez:2023nnl} have holographically investigated  the dynamics of a strongly coupled field theory on I-branes, specifically D5 branes intersecting on a line. An extension of our work could be finding some pulsating string solution on the background II mentioned in \cite{Nunez:2023nnl}. As another extension of this work, one can study the perturbation of these pulsating string solutions. We hope to come up with some of these investigations in near future. 

\section*{Acknowledgements} 
SB to thank Aritra Banerjee for some useful discussions and continuous encouragement throughout this work.

\end{document}